\documentclass[twocolumn,twoside,slac_two]{revtex4}
\usepackage{graphicx}
\usepackage{fancyhdr}
\usepackage{graphics}
\usepackage{epstopdf}
\usepackage{textpos}
\newcommand{\pip}{\pi^+}
\newcommand{\pim}{\pi^-}
\newcommand{\piz}{\pi^0}
\newcommand{\etap}{\eta^{\prime}}
\newcommand{\etac}{\eta_c}
\newcommand{\hc}{h_c}

\newcommand{\psp}{\psi^{\prime}}
\newcommand{\chicJ}{\chi_{cJ}}
\newcommand{\chicz}{\chi_{c0}}
\newcommand{\chico}{\chi_{c1}}
\newcommand{\chict}{\chi_{c2}}

\newcommand{\jpsi}{J/\psi}

\newcommand{\pp}{\pi^+\pi^-}
\newcommand{\kk}{K^+K^-}
\newcommand{\ks}{K_S}

\newcommand{\kskp}{K_SK^+\pi^-}
\newcommand{\kkpiz}{\kk\piz}
\newcommand{\kskppp}{K_SK^+\pp\pi^-}
\newcommand{\kkpppiz}{\kk\pp\piz}

\newcommand{\pppppp}{3(\pp)}

\newcommand{\etapp}{\eta\pp}

\newcommand{\jpsipp}{\pi^+\pi^- J/\psi}

\newcommand{\mev}{\mathrm{MeV}}

\newcommand{\mevcc}{\mathrm{MeV}/c^2}

\begin{document}

\title{\centering  Charmonium spectroscopy and decays}
\author{
\centering
\begin{center}
J. Zhang
\end{center}}
\affiliation{\centering IHEP, Beijing, 100081, P.R.China}
\begin{abstract}
In this talk, I review the recent experimental developments on
charmonium. These mainly include the precision measurements of
spin-singlet states $\etac$, $\etac(2S)$, $h_c$,
studies of the charmonium-like states $X(3872)$ and the $Y$ states. 
Charmonium transitions and decays are also discussed.
\end{abstract}

\maketitle
\thispagestyle{fancy}


\section*{Introduction}
\begin{figure}[pt]
  \includegraphics[width=80mm]{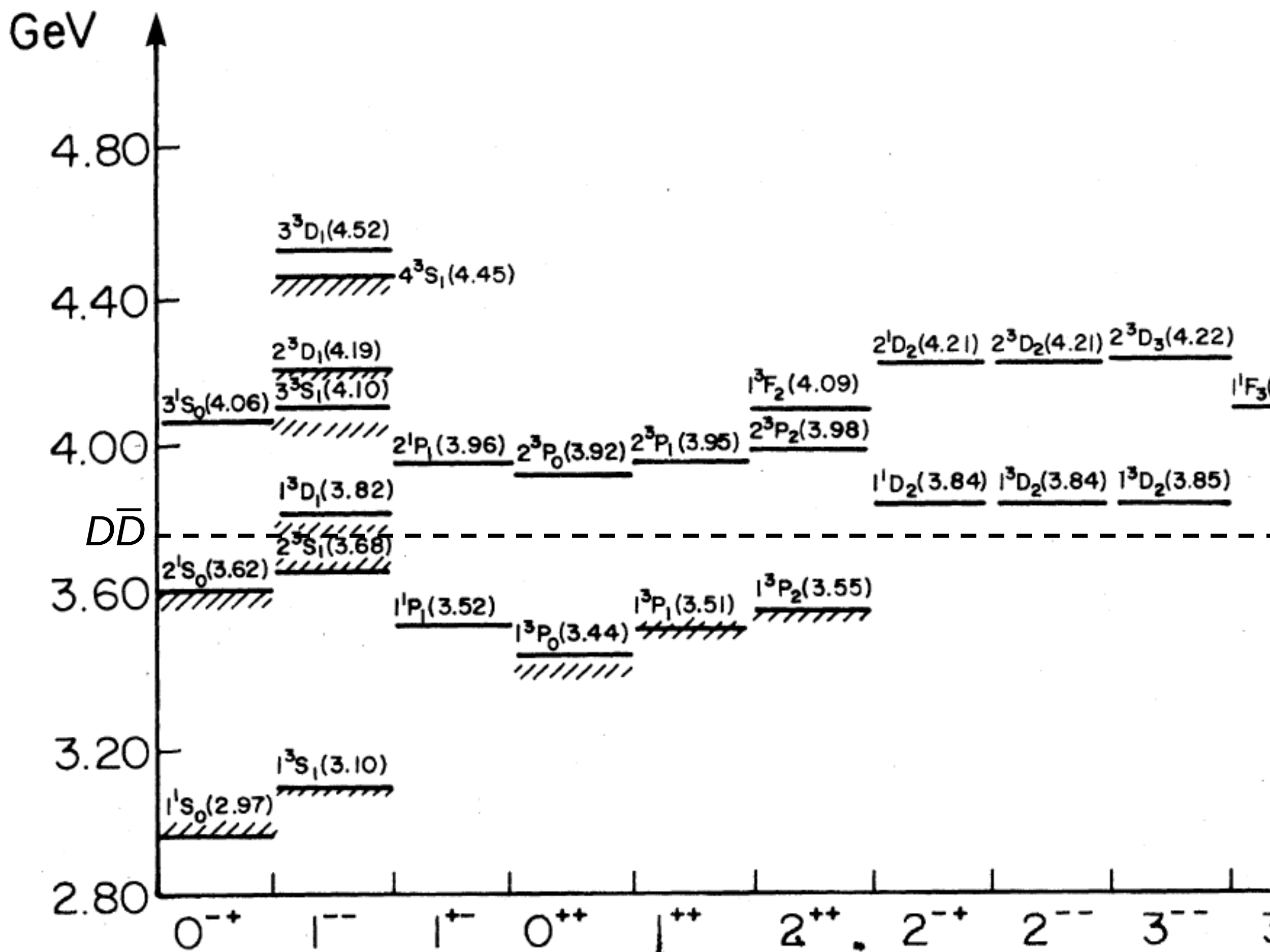}
  \vspace*{0.5cm}
  \caption{The charmonium states~\cite{Godfrey:1985xj}. Dashed line denotes the
    $D\bar{D}$ mass threshold.}
  \label{charmonium_spect}
\end{figure}

Charmonia are charmed-quark and anticharmed-quark states
($c\bar{c}$) bound by the strong interaction. Charmed quarks are
heavy, so the motion of the charm quark inside the bound state is slow, $v^2\sim
0.3$, where $v$ is relative velocity between the $c$ and $\bar{c}$. 
The charmonium system can be approximately considered as a
non-relativistic bound state. The energy levels can be found by
solving a non-relativistic Schrodinger equation, with sophisticated
corrections (e.g. relativistic correction) and other effects. 
Figure~\ref{charmonium_spect} shows the charmonium levels from this
approach~\cite{Godfrey:1985xj}.
Although all charmonium states below the $D\bar{D}$ mass threshold have been
observed, knowledge is sparse on spin-singlet $S$-wave
$\etac$($1^1S_0$), the $\etac(2S)$($2^1S_0$), and the $P$-wave
$h_c$($1^1P_1$).
Above the threshold, the spin-triplet $S$-wave states $\psi(4040)$,
and $\psi(4415)$, and the $D$-wave $\psi(3770)$, and $\psi(4160)$ have been found.
The $\psi(4040)$, $\psi(4415)$, $\psi(3770)$, $\psi(4160)$ are
commonly assigned as $\psi$($3^3S_1$), $\psi$($4^3S_1$),
$\psi$($1^3D_1$) and $\psi$($2^3D_1$), respectively.
The $Z(3930)$ observed by Belle~\cite{Uehara:2005qd} in the $D\bar{D}$
mass distribution from $e^+e^-\to e^+e^- D\bar{D}$ events, is
identified with $\chict'$ ($2^3P_2$).

\section*{The $\hc$ $(1^1P_1)$}
Information about the spin-dependent interaction of heavy quarks
can be obtained from precise measurement of the $1P$ hyperfine mass
splitting $\Delta~M_{hf}\equiv\langle M(1^3P)\rangle- M(1^1P_1)$,
where $\langle
M(1^3P_{J})\rangle=(M(\chi_{c0})+3M(\chi_{c1})+5M(\chi_{c2}))/9=
3525.30\pm0.04$~MeV/$c^2$~\cite{PDG} is the spin-weighted centroid of
the $^3P_J$ mass and $M(1^1P_1)$ is the mass of the singlet state
$\hc$. A non-zero hyperfine splitting may give an indication of non-vanishing
spin-spin interactions in charmonium potential models~\cite{swanson}.

\begin{figure}[bpt]
  \includegraphics[width=80mm]{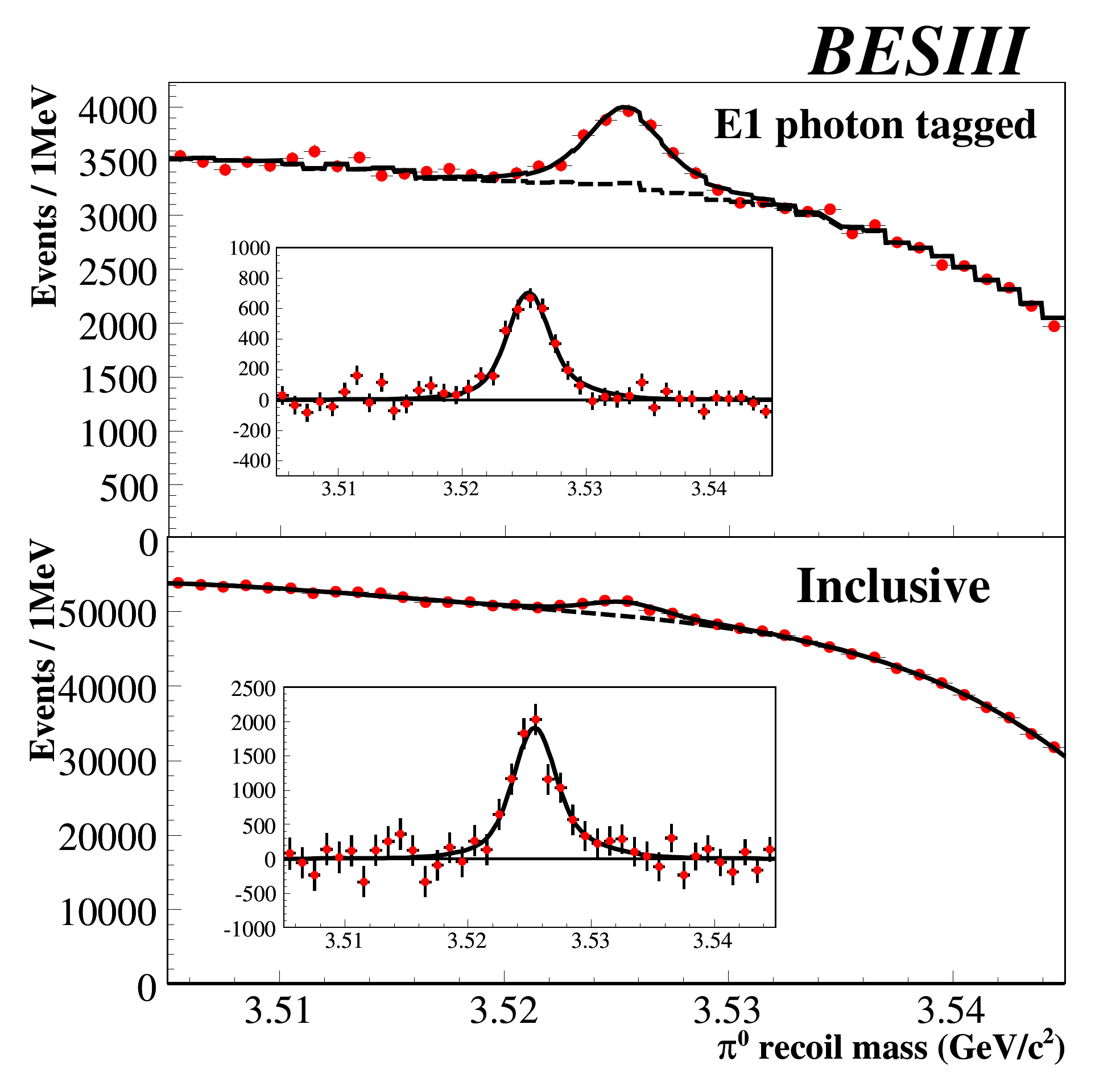}
  \caption{Top: the $\pi^0$ recoil mass spectrum and fit for the
    $E1$-tagged analysis of $\psi'\to \pi^0\hc,  \hc\to\gamma\eta_c$;
    bottom: the $\pi^0$ recoil mass spectrum and fit for the inclusive
    analysis of $\psi'\to\pi^{0}\hc$~\cite{Ablikim:2010rc}. Fits are shown as solid lines,
    background as dashed lines. The insets show the
    background-subtracted spectra.}
  \label{hc-FitData}
\end{figure}

With 106M $\psi'$ events, BESIII observed clear
signals in the $\piz$ recoil mass distribution for
$\psp\to\piz \hc$ with and without the subsequent radiative decay
$\hc\to\gamma\etac$~\cite{Ablikim:2010rc}, as shown in Fig.~\ref{hc-FitData}. 
They reported first measurements of the absolute branching ratios
$\mathcal{B}(\psp \rightarrow \piz \hc) = (8.4 \pm 1.3 \pm
1.0) \times 10^{-4}$ and $\mathcal{B}(\hc \rightarrow \gamma \etac) =
(54.3 \pm 6.7 \pm 5.2)\%$, along with improved measurements of the
$\hc$ mass $M(\hc) = 3525.40 \pm 0.13 \pm 0.18~\mevcc$. 
They found the $1P$ hyperfine mass splitting to be 
$\displaystyle \Delta~M_{hf} \equiv
\langle M(1^3P) \rangle - M(1^1P_1) =-0.10 \pm 0.13 \pm0.18~\mevcc$,
which is consistent with no strong spin-spin interaction.
The results are in agreement with CLEO-c's earlier
 results~\cite{Dobbs:2008ec}.

In addition, BESIII used 16 exclusive hadronic $\etac$ decay modes to reconstruct
$\hc\to \gamma\etac$. 
By doing so, the ratio of signal to background can be improved significantly. 
A simultaneous fit to the $\piz$ recoil mass distributions of the 16
decay modes was performed. 
From 106M $\psp$ events, $835\pm 35$ signal events are found. The
measured $\hc$ mass and width are
$m  =3525.31\pm 0.11 \pm0.15~\mevcc$, and 
$\Gamma =0.70\pm0.28\pm0.25~\mev$,
which are consistent with the inclusive analysis results, and also
consistent with CLEO-c's results~\cite{Dobbs:2008ec}.

\subsection*{The $\etac(2S)$}
The radially excited $n=2$ spin-singlet $S$-wave state, the $\etac(2S)$ meson,
was not well established until the Belle collaboration
found the $\etac(2S)$ signal at $3654\pm 6(stat)\pm
8(syst)$ $\mevcc$ in the $\kskp$ invariant mass distribution in a
sample of exclusive $\etac(2S)\to \kskp$ decays~\cite{Choi:2002na}. 
Since then measurements of $\etac(2S)$ in photon-photon fusion to
$K\bar{K}\pi$ final state have been
reported~\cite{Aubert:2003pt,Asner:2003wv,Nakazawa:2008zz}, as well as
in double charmonium production \cite{Aubert:2005tj,Abe:2007jn}.
CLEO-c searched for $\etac(2S)$ in the radiative decay
$\psp\to\gamma\etac(2S)$, found no clear signals in its sample of 25M
$\psp$~\cite{:2009vg}. The challenge of this measurement is the detection of 50 $\mev$ photons.

With 519 fb$^{-1}$, BaBar observed
$\etac(2S)\to\kskp$ and $\etac(2S)\to \kkpppiz$ produced in
photon-photon fusion for the first time~\cite{delAmoSanchez:2011bt}.
They measured the mass and width of $\etac$ and $\etac(2S)$ in $\kskp$
decays, $m(\etac(1S))= 2982.5 \pm 0.4 \pm 1.4~\mevcc$,
$\Gamma(\etac(1S)) = 32.1 \pm 1.1 \pm 1.3~\mev$,
$m(\etac(2S))= 3638.5 \pm 1.5 \pm 0.8~\mevcc$, $\Gamma(\etac(2S)) = 13.4 \pm 4.6 \pm 3.2~\mev$.
These $\etac(2S)$ results are so far the most precise measurements.

Belle updated the analysis of $B^\pm\to K^{\pm}\etac$ and $B^{\pm} \to
K^\pm \etac(2S)$ followed by $\etac$ and $\etac(2S)$ decay to $\kskp$
with 535 million $B\bar{B}$-meson pairs~\cite{:2011dy}.
Both decay channels contain the backgrounds from $B^\pm \to K^\pm\kskp$ decays
without intermediate charmonia, which could interfere with the
signal. Belle's analysis took interference into account with no
assumptions on the phase or absolute value of the interference. 
A two dimensional $M(\kskp)$ -- $\cos\theta$ fit was performed to extract signal,
where $\theta$ is the angle between $K$ (from $B$ directly) with
respect to $\ks$ in the rest frame of the $\kskp$.
They obtained the masses and widths of $\eta_c$ and $\eta_c(2S)$.
For the $\eta_c$ meson parameters the model error is negligibly small:
$M(\eta_c)=2985.4\pm1.5(stat)^{+0.2}_{-2.0}(syst)$ MeV/$c^2$,
$\Gamma(\eta_c)=35.1\pm 3.1(stat)^{+1.0}_{-1.6}(syst)$ MeV/$c^2$.
For the $\eta_c(2S)$ meson the model and statistical uncertainties cannot
be separated:
$M(\eta_c(2S))=3636.1^{+3.9}_{-4.1}(stat+model)^{+0.5}_{-2.0}(syst)$ MeV/$c^2$,
$\Gamma(\eta_c(2S))=6.6^{+8.4}_{-5.1}(stat+model)^{+2.6}_{-0.9}(syst)$ 
MeV/$c^2$.

\begin{figure}[bpt]
  \includegraphics[width=80mm]{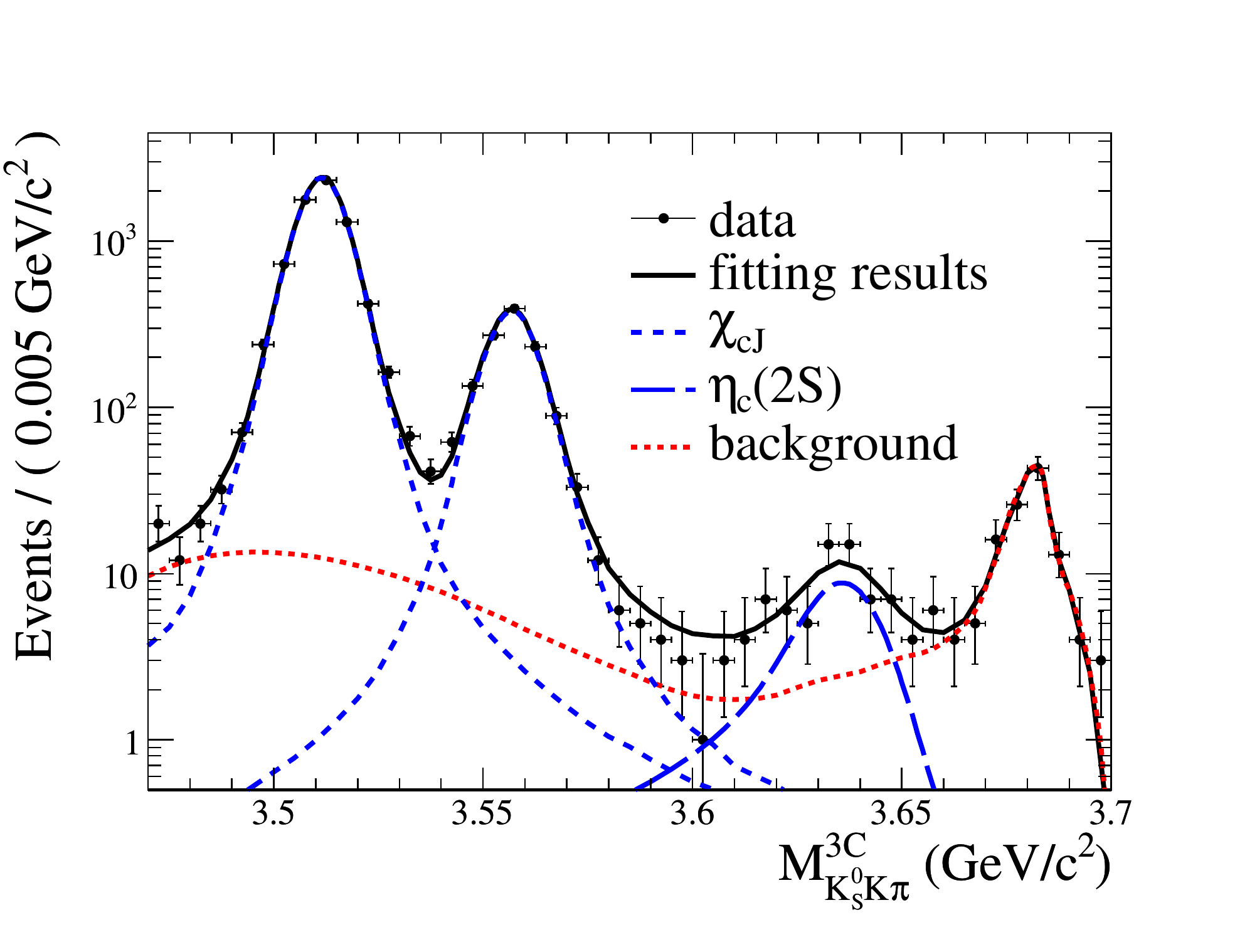}
  \caption{The $\kskp$ invariant mass for selected events for $\psp\to
  \gamma\kskp$. Points are data and the solid curve is the fit
  results. Blue long-dashed line is signal. Blue dashed lines are
  $\chico/\chict \to \gamma\kskp$ events. Red dotted line is for other
  backgrounds mainly from the decays $\psp \to \piz\kskp$,
  $\kskp$ and ISR/FSR production of $\kskp\gamma_{ISR}/\gamma_{FSR}$.}
  \label{fit_etacp}
\end{figure}

Using 106 million $\psp$ events, BESIII searched for $\etac(2S)$
in the decay $\psp\to \gamma\etac(2S)$, with  $\etac(2S)\to \kskp$.
Figure \ref{fit_etacp} shows the invariant mass distribution of
$\kskp$, where a three-constraints kinematic fit has been applied (in which
the energy of the photon is allowed to float). The solid curve
in Fig.~\ref{fit_etacp} shows preliminary results of an
unbinned maximum likelihood fit with four components: signal,
$\chi_{c1}$, $\chi_{c2}$ and other background (coming from $\psp$
decays to $\piz\kskp$, $\kskp$ and ISR/FSR production of $\kskp\gamma_{ISR}/\gamma_{FSR}$).
The fit, in which the width of the $\etac(2S)$ is fixed at 12 MeV, yields
$50.6\pm9.7$ signal events, and gives the mass $M(\etac(2S)) =3638.5 \pm2.3\pm1.0~\mevcc$.
The statistical significance of the signal is more than 6$\sigma$.
Using the detection efficiency determined from MC simulation, the
product branching fraction is obtained 
${\cal B} (\psp\to\gamma\etac(2S)) \times {\cal B}(\etac(2S)\to \kskp)
=(2.98\pm 0.57\pm0.48)\times 10^{-6}$.
Using the result ${\cal B}(\etac(2S)\to K\bar{K}\pi) =(1.9\pm0.4\pm
1.1)\%$ from BaBar~\cite{Aubert:2008kp} gives the branching fraction
${\cal B}(\psp\to\gamma\etac(2S)) =(4.7\pm0.9\pm 3.0)\times 10^{-4}$.
This result is consistent with CLEO-c's upper limit \cite{:2009vg} and
predictions of potential models \cite{Eichten:2002qv}.

\section*{The $\etac(1S)$} 
The mass and width of the lowest lying charmonium state, the $\etac$ ($1^1S_0$),
continue to have large uncertainties when compared to those of other
charmonium states~\cite{PDG}. Early measurements of the properties of
the $\etac$ using $\jpsi$ radiative transitions~\cite{Baltrusaitis:1985mr,Bai:2003et} found a mass and
width of $2978~\mevcc$ and $10~\mev$, respectively. However, recent
experiments, including photon-photon fusion and $B$ decays, have
reported a significantly higher mass and a much wider
width~\cite{Aubert:2003pt, Asner:2003wv,  Uehara:2007vb, belle2011}. 
The most recent study by the CLEO-c experiment,
using both $\psp \to \gamma\etac$ and $\jpsi\to \gamma\etac$, pointed
out a distortion of the $\etac$ line shape in $\psp$ decays~\cite{recent:2008fb}. CLEO-c
attributed the $\etac$ line-shape distortion to the energy dependence
of the ``hindered'' $M1$ transition matrix element.

\begin{figure}[bpt]
  \includegraphics[width=40mm]{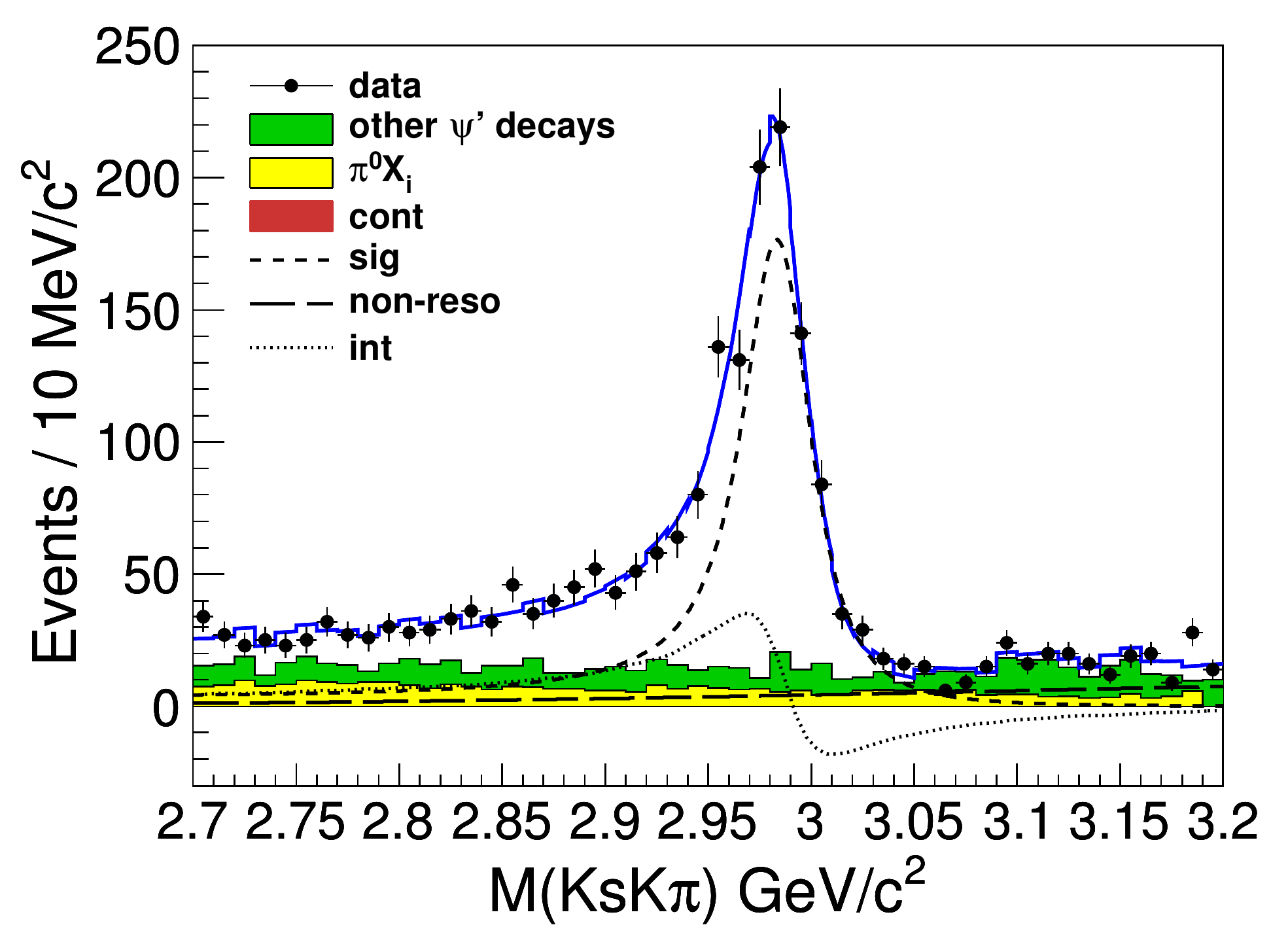}
  \includegraphics[width=40mm]{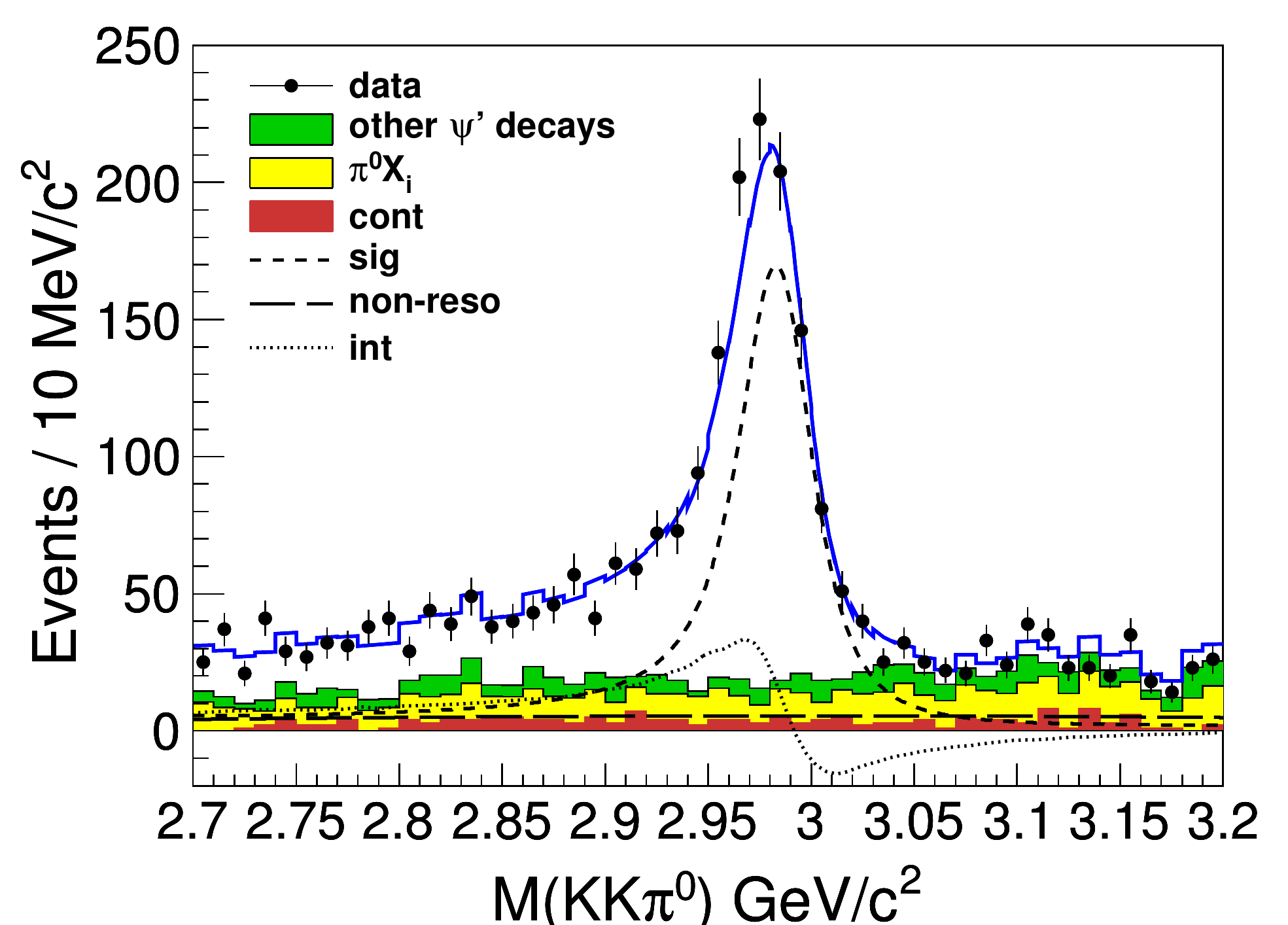}
  \includegraphics[width=40mm]{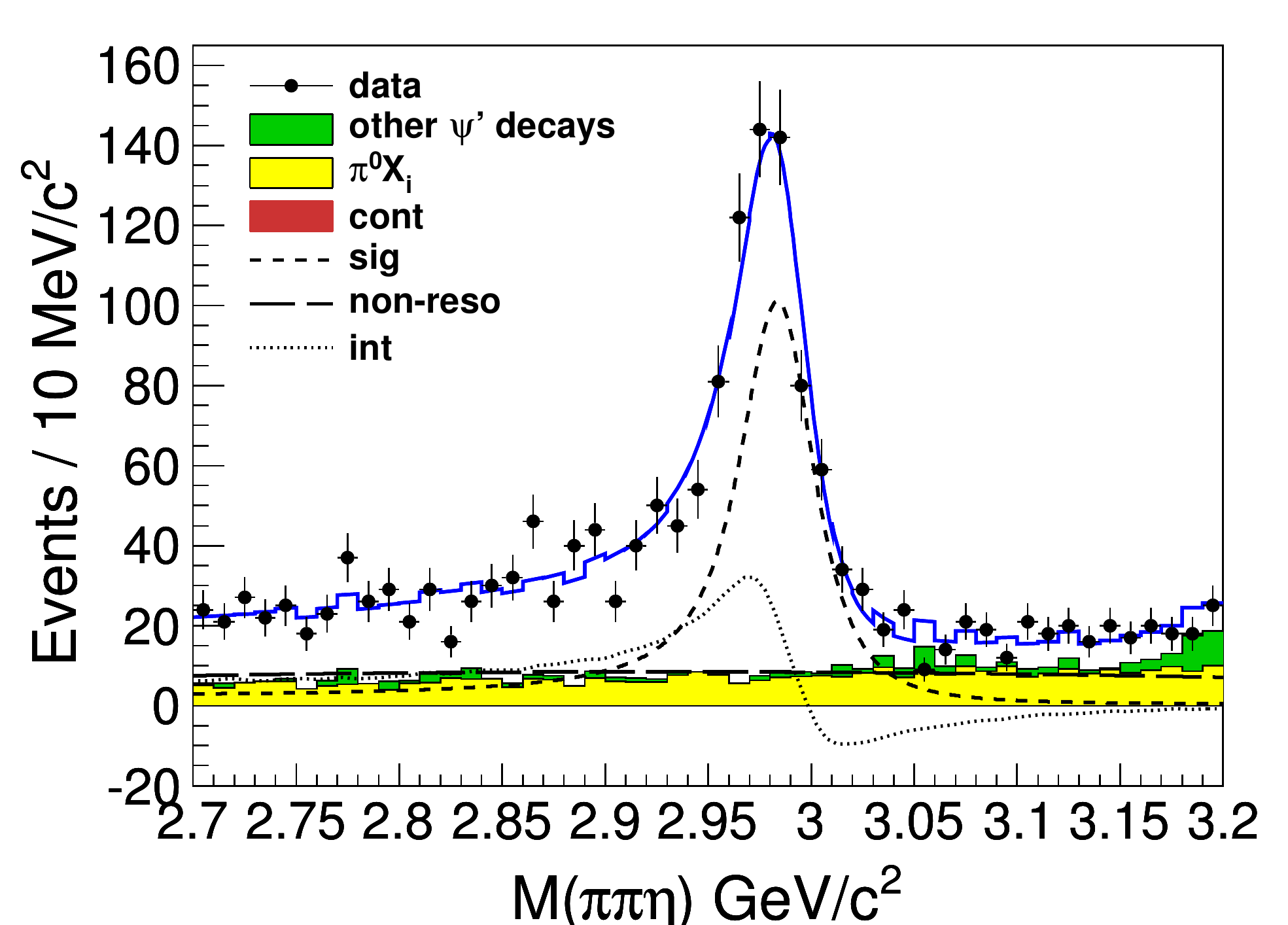}
  \includegraphics[width=40mm]{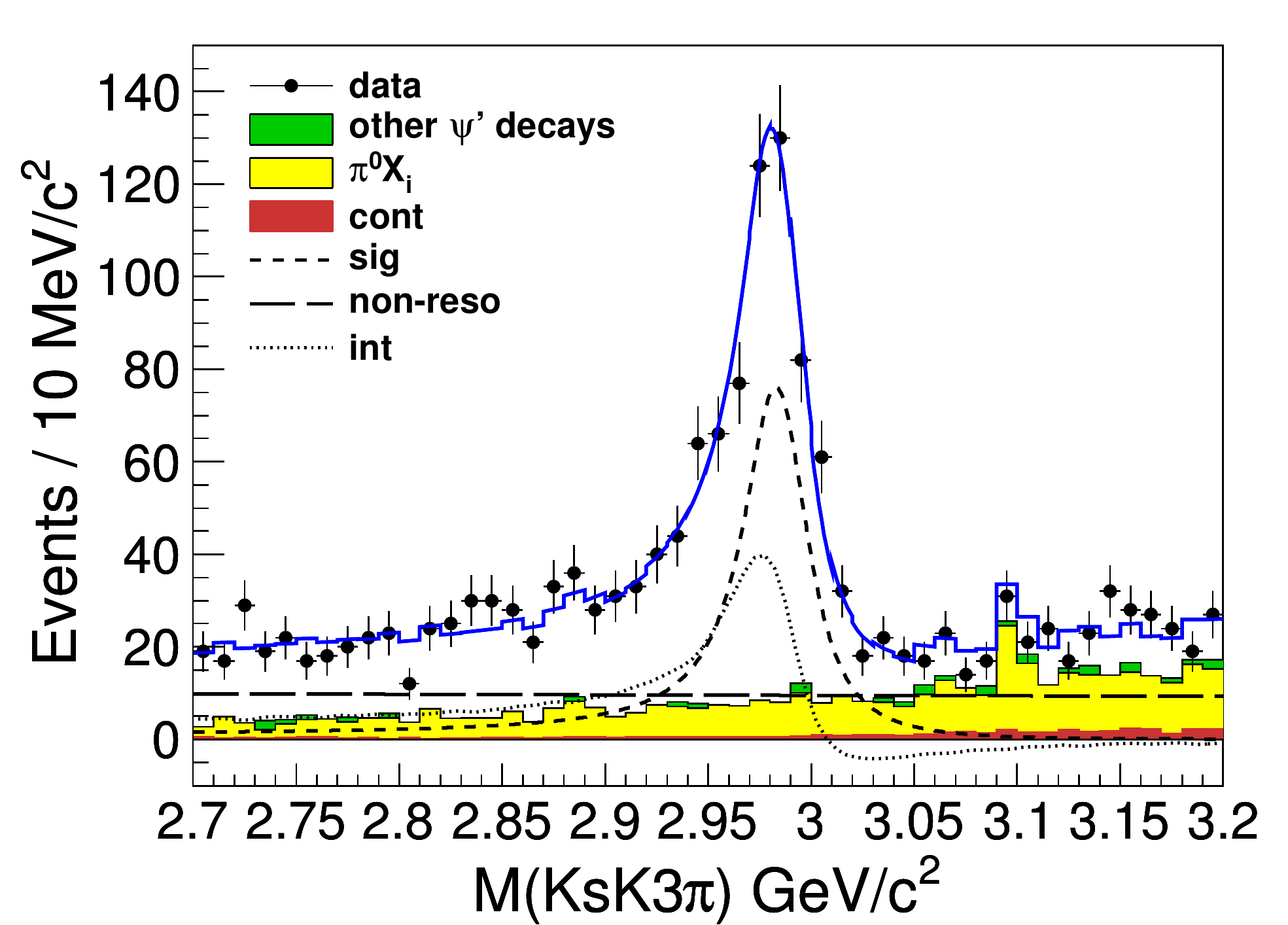}
  \includegraphics[width=40mm]{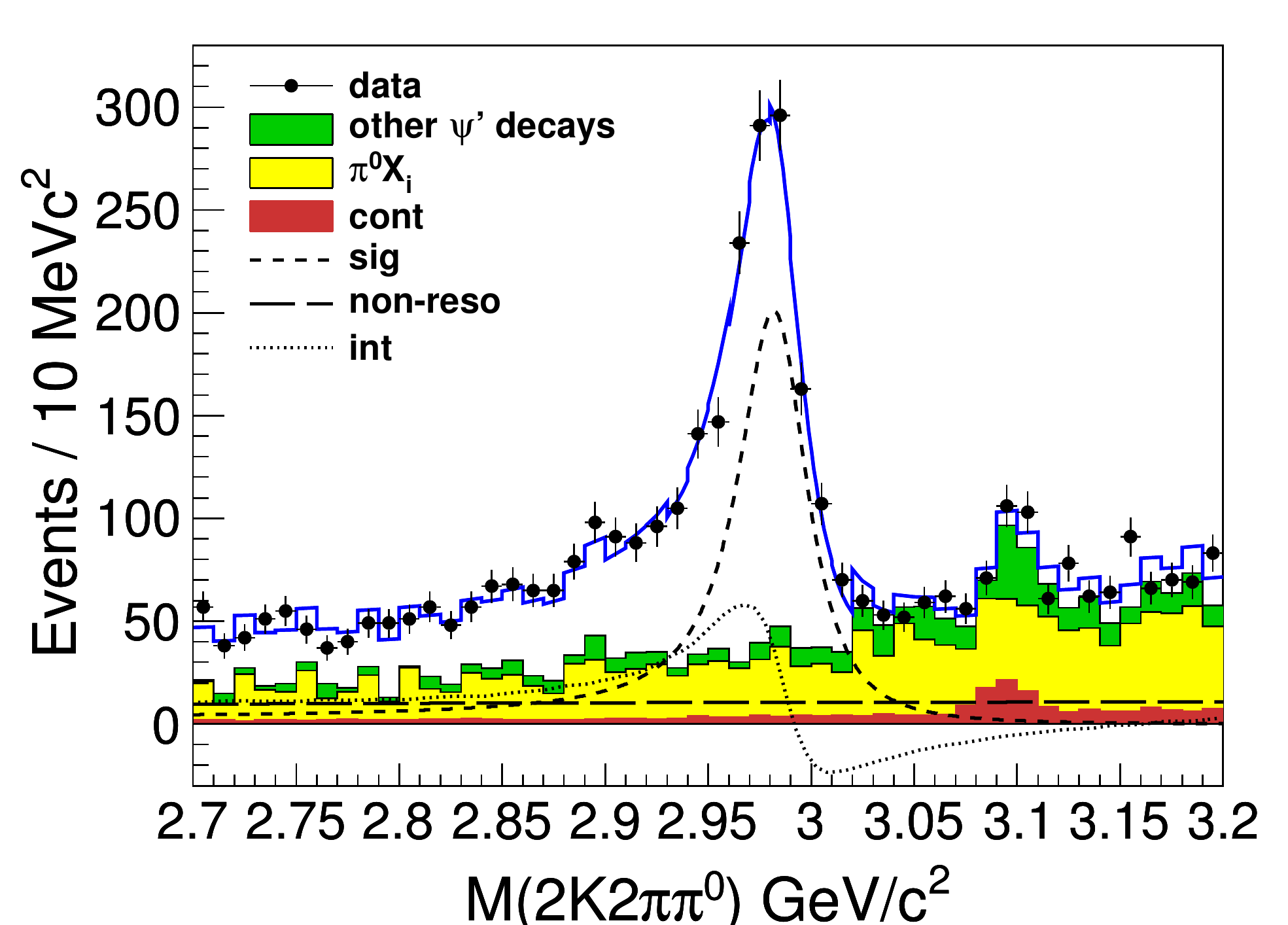}
  \includegraphics[width=40mm]{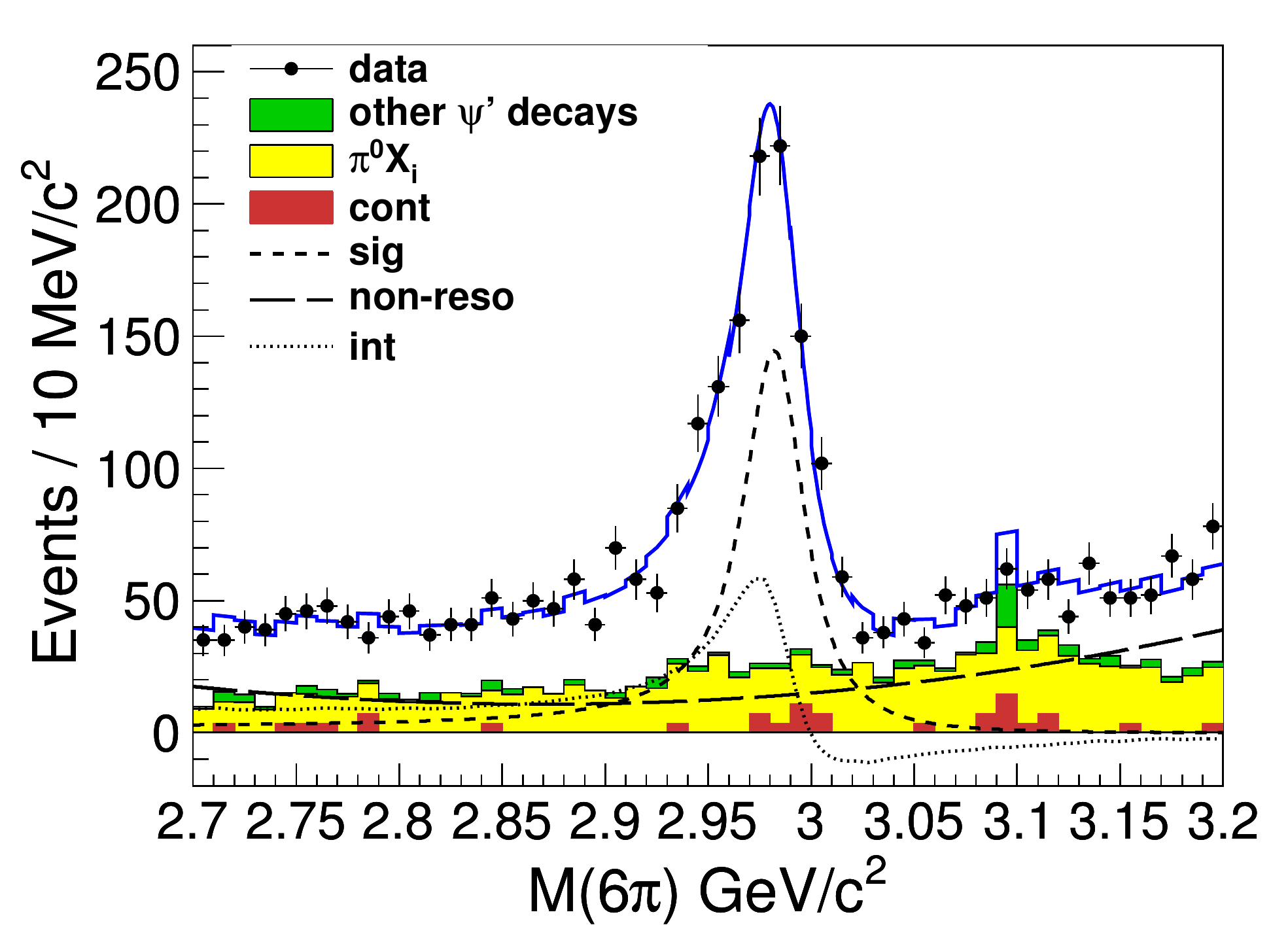}
  \caption{The $M(X_i)$ invariant mass distributions for the $\etac$
  decays to  $\kskp$, $\kkpiz$, $\etapp$, $\kskppp$, $\kkpppiz$ and $\pppppp$,
  respectively, with the fit results superimposed. Points are data and
  the solid lines are the total fit results. Signals are shown as
  short-dashed lines; the non-resonant components as long-dashed
  lines; and the interference between them as dotted lines.
  Shaded histograms are (in red/yellow/green) for (continuum/$\pi^0
  X_i$/other $\psp$ decays) backgrounds. The continuum backgrounds for
  $\kskp$ and $\etapp$ decays are negligible.}
  \label{fig:metac}
\end{figure}

At BESIII, the $\etac$ can be produced through $\psp \to \gamma\etac$,
and the $\etac$ mass and width are determined by fits to the
invariant mass spectra of exclusive $\etac$ decay modes. 
Six modes are used to reconstruct the $\etac$: $\kskp$, $\kkpiz$, $\etapp$,
$\kskppp$, $\kkpppiz$, and $\pppppp$, where the $\ks$ is reconstructed
in $\pp$, and the $\eta$ and $\piz$ in $\gamma\gamma$ decays.
Figure~\ref{fig:metac} shows the $\etac$ invariant mass
distributions for selected $\etac$ candidates, together with the
estimated $\piz X_i$ backgrounds ($X_i$ represents the $\etac$ final
states under study), the continuum backgrounds normalized by
luminosity, and other $\psp$ decay backgrounds estimated from the
inclusive MC sample.
A clear $\etac$ signal is evident in every decay mode. The $\etac$
signal has an obviously asymmetric shape that suggests possible
interference with a non-resonant $\gamma X_i$ amplitude.
%
The fitted relative phases between the signal and the non-resonant component
from each mode are consistent within $3\sigma$,
which may suggest a common phase in all the modes under study. A fit
with a common phase (i.e. the phases are constrained to be the same)
describes the data well.
The preliminary results on the $\etac$ mass and width are
$M    = 2984.4\pm  0.5(stat.)\pm 0.6(syst.)~\mevcc$,
$\Gamma = 30.5\pm  1.0(stat.)\pm 0.9(syst.)~\mev$.

\begin{figure}[bpt]
  \includegraphics[width=40mm]{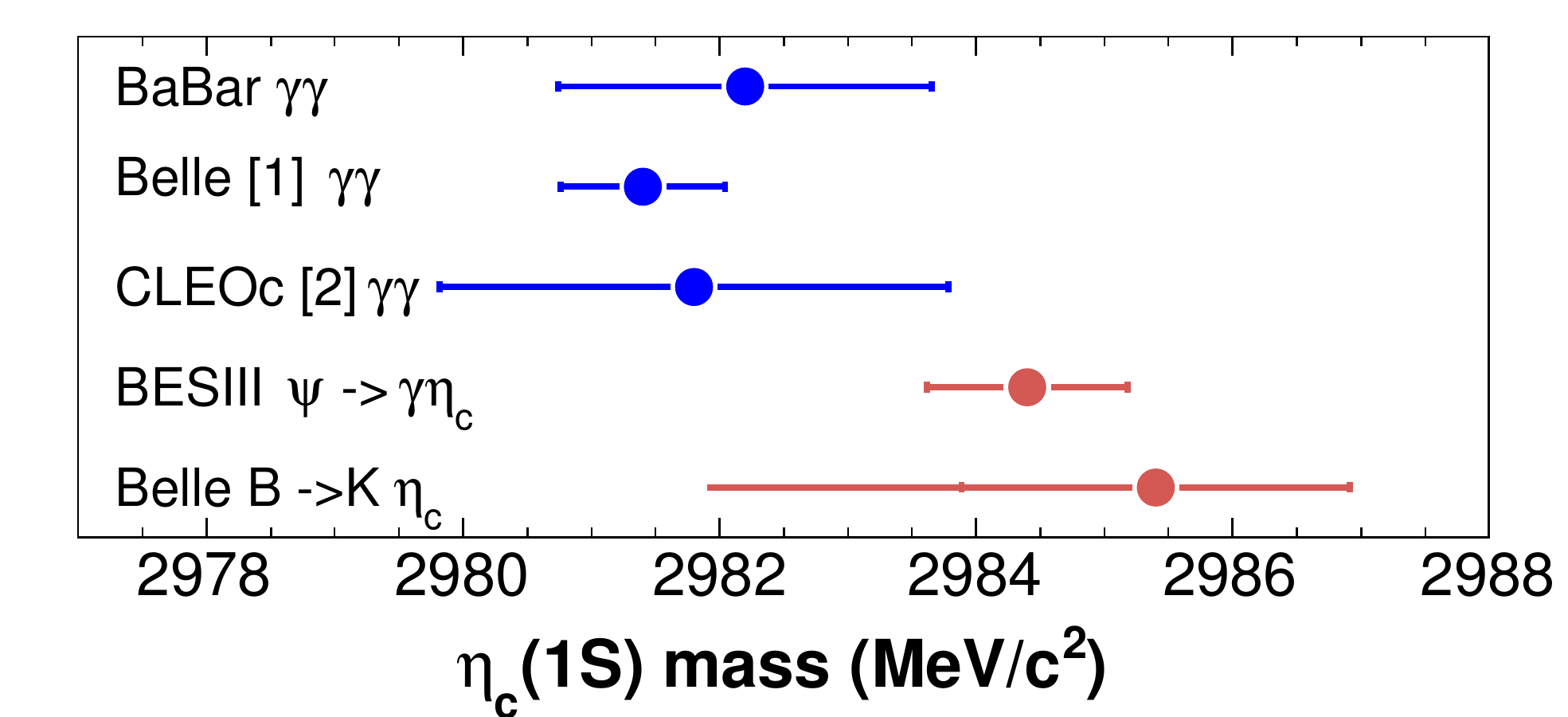}
  \includegraphics[width=40mm]{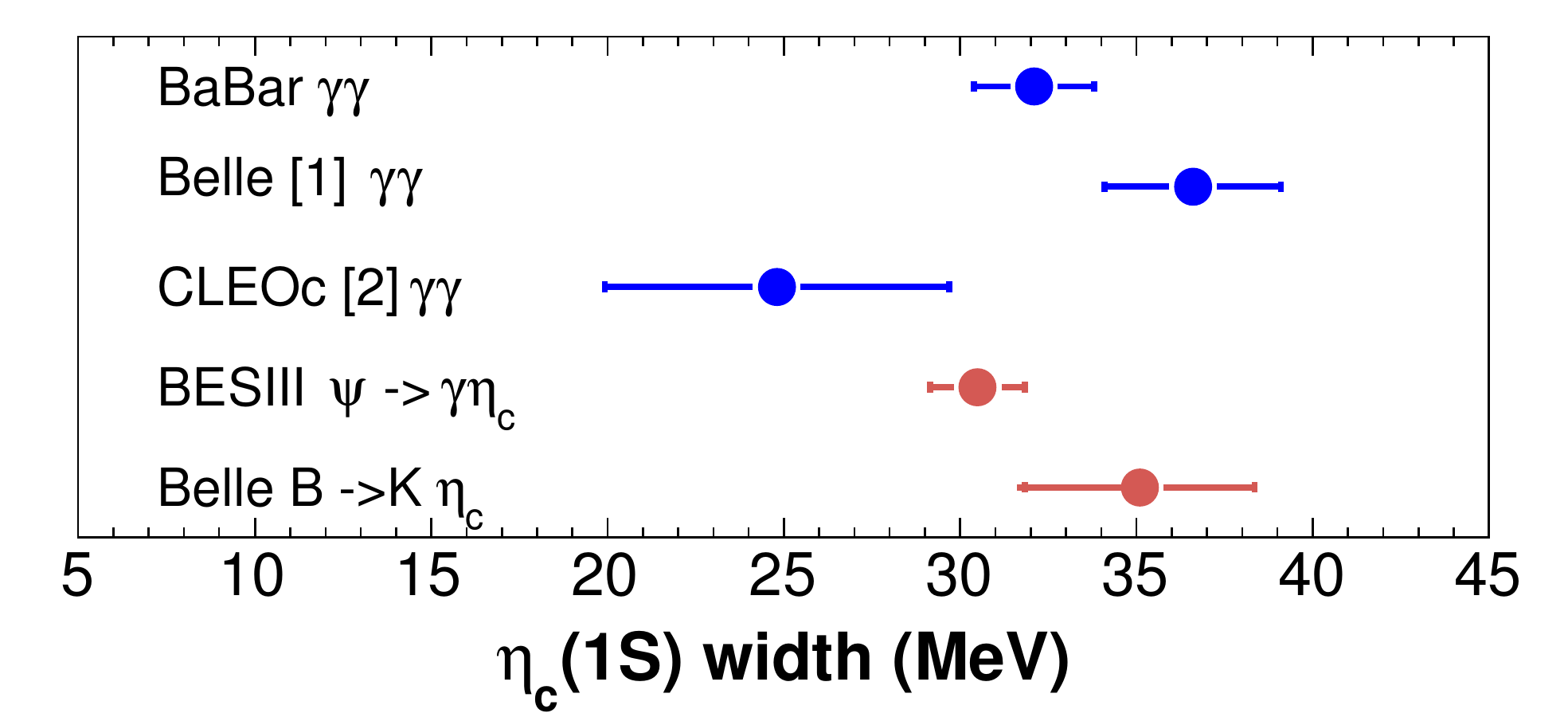}
  \includegraphics[width=40mm]{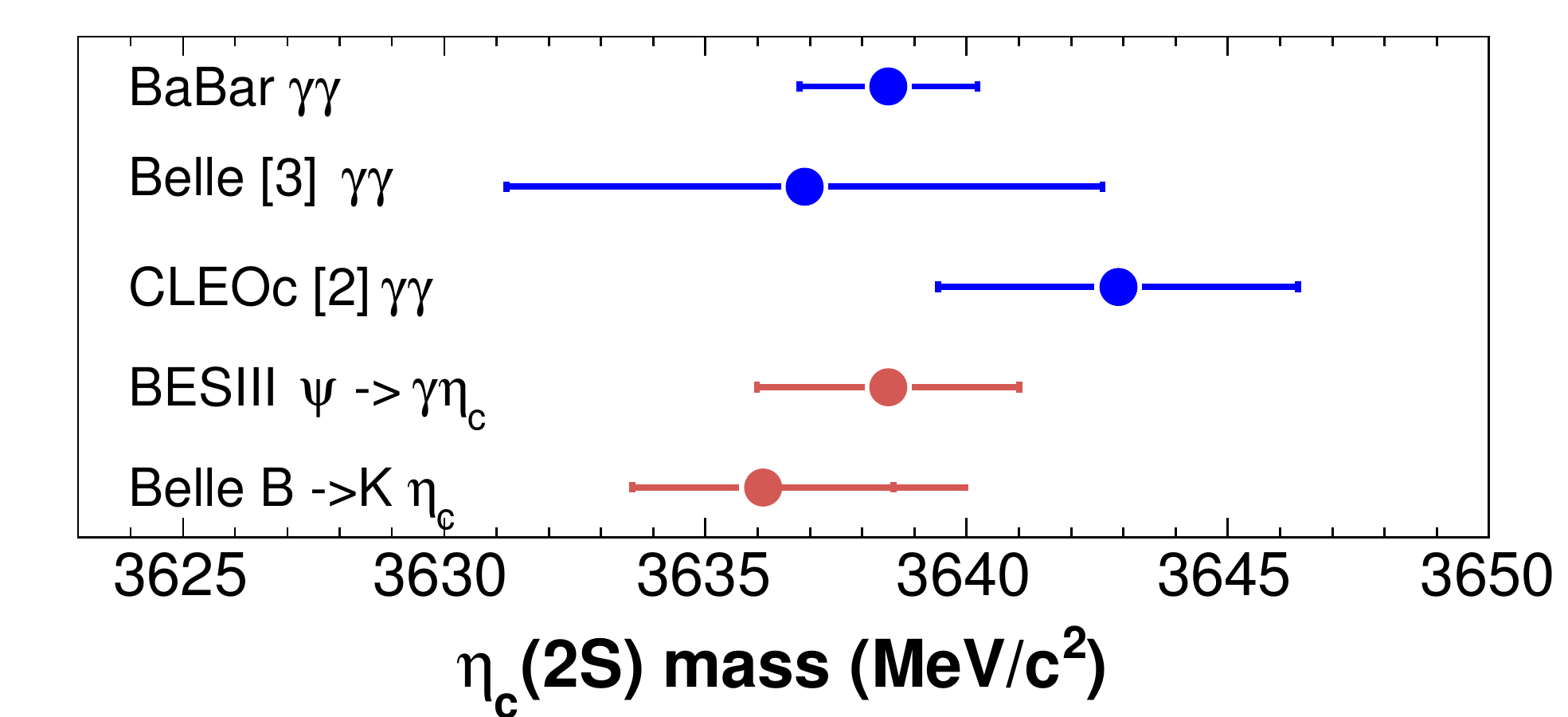}
  \includegraphics[width=40mm]{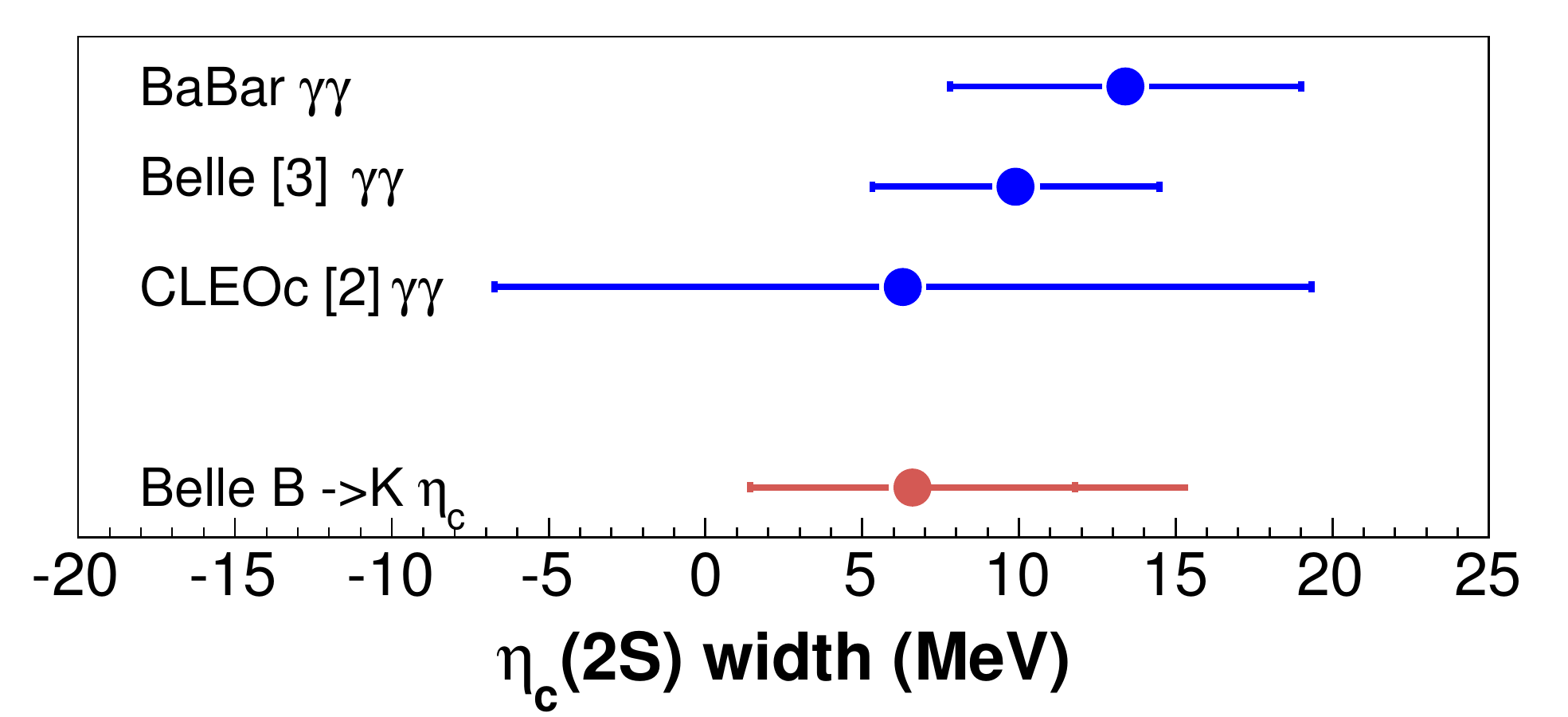}
  \caption{Comparison of the masses (left) and widths (right)
    of the $\etac$ (top) and $\etac(2S)$ (bottom) measured from from
    the two-photon process, $\psp$ transition, and $B$ decays.
  }
  \label{comparison_etac}
\end{figure}

Figure~\ref{comparison_etac} compares the recent
measurements of the $\etac$ and $\etac(2S)$ mass and width from two
photon-photon fusion, $\psp$ transition, and $B$ decays. These results
are in good agreement.
Hyperfine splittings are 
$\Delta M(1S) = 112.5\pm 0.8 \mev$, and
$\Delta M(2S) = 47.6\pm 1.7 \mev$.
which agree well with recent lattice computations~\cite{Burch:2009az}.

\section*{Charmonium-like states}
The $X(3872)$ was first observed as a narrow peak in the $\jpsipp$
invariant mass spectrum near $D^{*0}\bar{D^0}$ threshold from $B\to
K\jpsipp$ decays by Belle~\cite{Choi:2003ue} in 2003, and later
confirmed by BaBar~\cite{Aubert:2008gu}. 
The $X(3872)\to \jpsipp$ decay was also observed inclusively in prompt
production from $p\bar{p}$ collisions at the Tevatron by both CDF~\cite{Acosta:2003zx} and
D0~\cite{Abazov:2004kp}. CDF studied the angular distributions and correlations of the
$\jpsipp$ final state, found that the dipion was favored to originate
from $\rho^0 \to \pp$, and thus only $J^{PC}$ assignments of $1^{++}$ and
$2^{-+}$ explained their measurements~\cite{Abulencia:2006ma}.  

A number of theoretical models have been proposed for the $X(3872)$
states, such as conventional charmonium state, $DD^*$ molecules,
diquark-diantiquarks, $cc$-gluon hybrids etc., but none can
comfortably account for all experimental results. 
Charmonium states
$1^{++}$ $\chi_{c1}(2P)$ and $2^{-+}$ $\eta_{c2}(^1D_2)$ are possible
candidate states for the $X(3872)$. 
For a $\chi_{c1}(2P)$ state, a large $\chi_{c1}(2P)\to \gamma \jpsi$
branching fraction is expected; experimental results do not agree.
For a $\eta_{c2}(^1D_2)$ state, a large width is expected; but the observed
$X(3872)$ is narrow. In the $DD^*$ molecule model it is hard to explain the large radiative decay
rate, the $\pi\pi\jpsi$ rate and the production in $p\bar{p}$.
The diquark-diantiquarks model predicts partners for the
$X(3872)$, but no partner has been found yet. For $cc$-gluon hybrids, the
mass is too low. 

\section*{The mass of $X(3872)$}
\begin{figure}[bpt]
  \includegraphics[width=70mm]{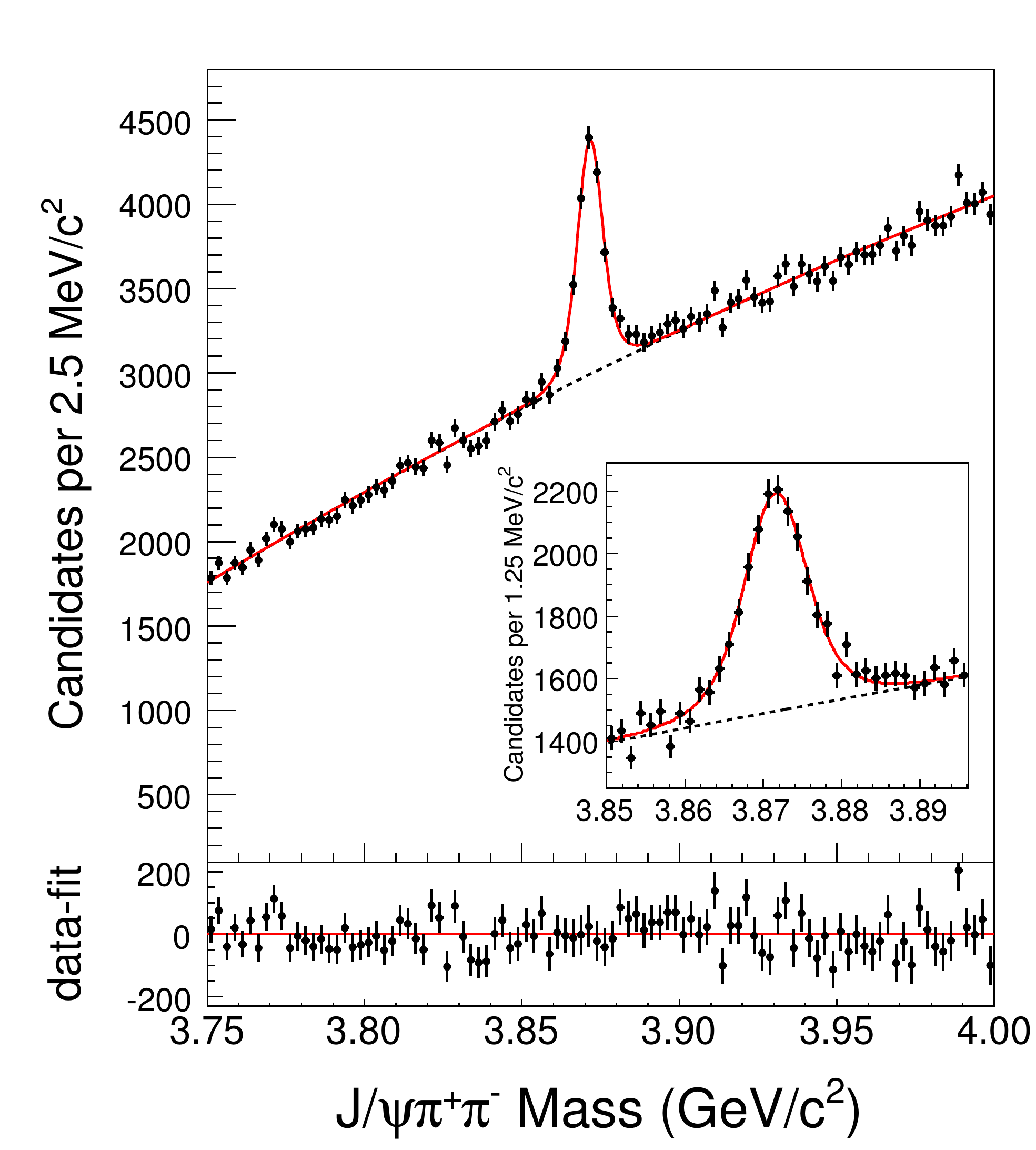}
  \caption{Invariant mass distribution of the $X(3872)$ candidates.
    The points show the data distribution, the full line is the projection of the
    unbinned maximum-likelihood fit, and the dashed line
    corresponds to the background part of the fit.
    The inset shows an enlargement of the region around the $X(3872)$ peak.
    Residuals of the data with respect to the fit are displayed below the mass plot.}
  \label{fig:CDF_massDistribution}
\end{figure}

With 2.4 fb$^{-1}$data, CDF presented an analysis of the mass of the
$X(3872)$ reconstructed via its decay to $\pi\pi\jpsi$~\cite{Aaltonen:2009vj}. They found
$\sim$ 6K candidates, shown in Fig.~\ref{fig:CDF_massDistribution}.  
They measured the $X(3872)$ mass 
$M_X = 3871.61\pm 0.16 (stat)\pm 0.19 (syst)~\mevcc$, 
which is the most precise determination to date.
In EPS2011, LHCb presented  measurements of the $X(3872)$ mass of
$M_X=3871.97 \pm 0.46\pm 0.1$ with 35 pb$^{-1}$ data.
Belle also updated the mass and width measurements with 711~fb$^{-1}$
data~\cite{Choi:2011fc}. A new world average that includes these new
measurements and other results that use the $\pi\pi\jpsi$ decay mode
is $M_X = 3871.67\pm 0.17~\mevcc$.

\begin{figure}[bpt]
  \includegraphics[width=70mm]{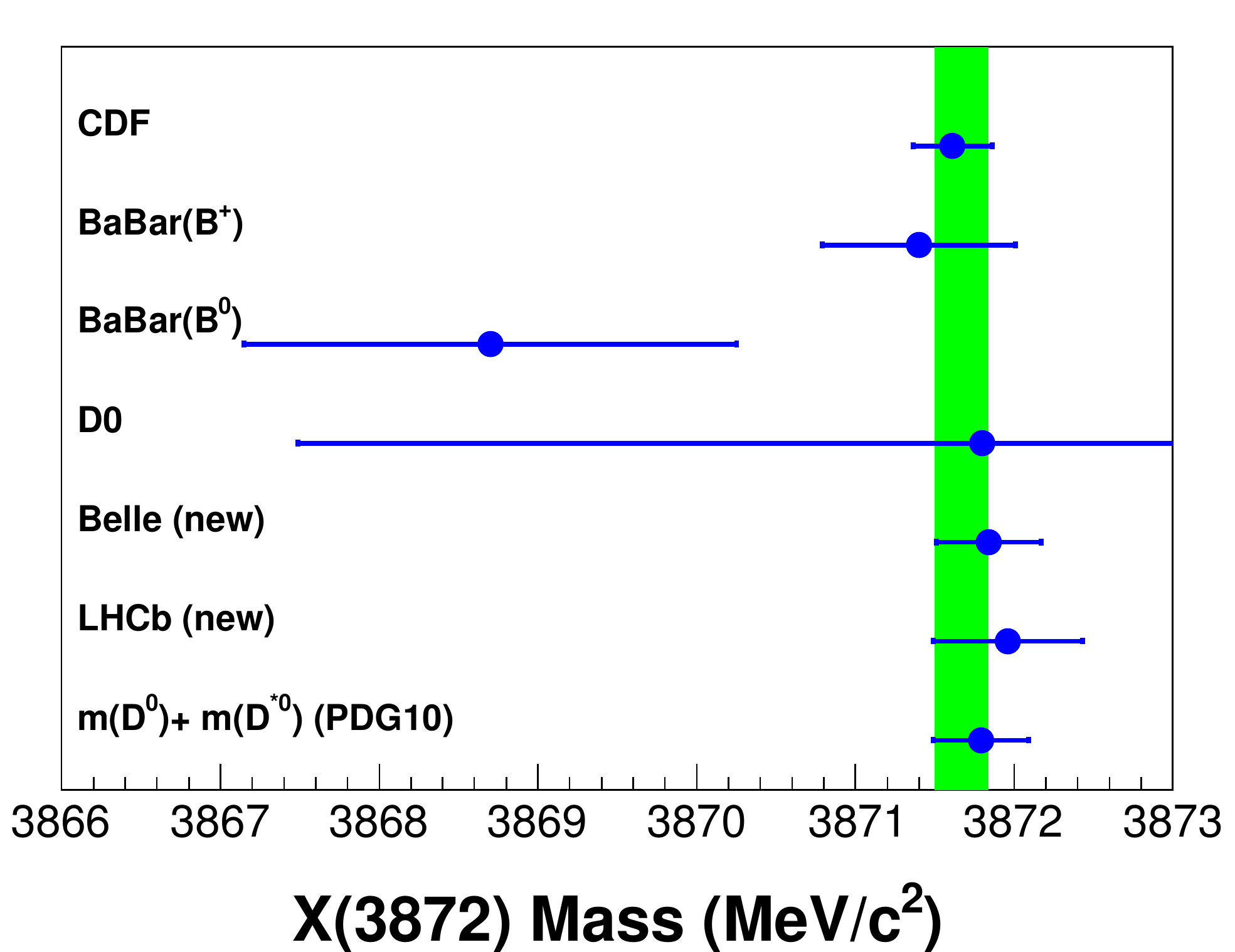}
   \caption{Comparison of the mass measurements of the $X(3827)$.}
  \label{compare_x_mass}
\end{figure}

An important feature of the $X(3872)$ is its mass is close to the
$\bar{D}^0D^{*0}$ threshold. A possible interpretation is that the $X(3872)$
is a molecule-like arrangement comprised of a $D^{*0}$ and a $\bar
D^0$~\cite{Voloshin:2003nt,Tornqvist:2004qy}. Crucial to these models
is whether the $X(3872)$ mass is above or below $m(D^{*0})+m(\bar D^0)$. 
Taking the $D^0$ and $D^{*0}$ mass
from PDG 2010, $m(D^{*0})+m(\bar{D}^0) = 3871.79\pm0.30~\mevcc$. The
new world average is $0.12\pm 0.35~\mevcc$  lower than 
$m(D^{*0})+m(\bar D^0)$, as shown in Fig.~\ref{compare_x_mass}.

\section*{Search for $X(3872)$ partners} 
Models of tightly bound diquark-antidiquark system feature two neutral
and one charged partner states~\cite{Maiani:2004vq}. The $X(3872)$,
observed in $B^+$ decays, is interpreted as a $cu\bar{c}\bar{u}$ combination. 
In $B^0\to K_S\jpsipp$, one should see a partner state, the
$cd\bar{c}\bar{d}$ combination. The two states differ in mass by a few
$\mev$. 
In addition, a neutral $cd\bar{c}\bar{s}$ partner and a charged ($cu\bar{c}\bar{d}$)
partner are also expected by isospin and flavor-SU(3).

BaBar~\cite{Aubert:2008gu} and Belle~\cite{Choi:2011fc} measured
the $X(3872)$ mass for 
$B^+\to K^+\pi^+\pi^- \jpsi$ and $B^0\to K_S\pi^+\pi^- \jpsi$
decays, separately. They found mass differences that are consistent with
zero, $\Delta M_x= -0.69\pm 0.97\pm0.19~\mevcc$ for Belle, and 
$\Delta M_x = 2.7\pm 1.6\pm 0.4\mevcc$ for BaBar.
The possibility that the $X(3872)$ enhancement in $\jpsipp$ is
composed of two different narrow states, $X_L$ and $X_H$, was
addressed by CDF~\cite{Aaltonen:2009vj}. 
By fitting their $\sim$ 6000 event $X(3872)\to\jpsipp$ peak with
two different Gaussian functions, they found $X_L$ and $X_H$ have masses closer
than 3.6 MeV for equal $X_L$ and $X_H$ production.

BaBar searched for a charged partner of the $X(3872)$ in the
$\pi^+\piz \jpsi$ mass distribution from $B\to K\pi^+\piz \jpsi$
decays, and found no evidence for a signal in either $B^0$ or $B^+$
decays~\cite{Aubert:2004zr}. They determined  the upper limits 
${\cal B}(B^0\to K^-X^+)\times {\cal B}(X^+\to\rho^+\jpsi)<5.4\times 10^{-6}$, 
${\cal B}(B^+\to K^0X^+)\times {\cal B}(X^+\to\rho^+\jpsi)<22 \times
10^{-6}$ at 90\% CL. 
The Belle limits for the same quantities at 90\% CL are
${\cal B}(B^0\to K^-X^+)\times {\cal B}(X^+\to\rho^+\jpsi)<4.2\times 10^{-6}$, 
${\cal B}(B^+\to K^0X^+)\times {\cal B}(X^+\to\rho^+\jpsi)<6.1\times 10^{-6}$~\cite{Choi:2011fc}.
The results rule out the isospin triplet model for the $X(3872)$.

\section*{The $X(3872)\to \gamma \jpsi(\psp)$}
Using a data sample of 465 million $B\bar{B}$ pairs, BaBar searched for
$B\to c\bar{c}\gamma K$ decays, found evidence for $X(3872)\to
\jpsi\gamma$ and X(3872) $\to \psp\gamma$ with 3.6~$\sigma$ and
3.5~$\sigma$, respectively~\cite{:2008rn}.
They measured the product of branching fractions 
${\cal B}(B^\pm\to X(3872)K^\pm)\cdot {\cal B}(X(3872)\to
\jpsi\gamma) =(2.8\pm 0.8 \pm 0.1)\times 10^{-5}$ and 
${\cal B}(B^\pm\to X(3872)K^\pm)\cdot {\cal B}(X(3872)\to
\psp\gamma) =(9.5 \pm 2.7 \pm 0.6)\times 10^{-6}$, and obtained the ratio
$\frac{{\cal B}(X(3872)\to \psp\gamma)}{{\cal B}(X(3872)\to
  \jpsi\gamma)}= 3.4\pm 1.4$.
The relatively large branching fraction of $X(3872)\to \psp\gamma$ is
generally inconsistent with a purely $\bar{D}^0 D^{*0}$ molecular
interpretation of the $X(3872)$, and possibly indicates mixing with a
significant $c\bar{c}$ component.

With 772 million $B\bar{B}$ events, Belle
observed $X(3872)\to \jpsi
\gamma $ in the charged decay $B^+ \to X(3872)K^+$ with a
significance of $4.9\sigma$, while in a search
for $X(3872)\to \gamma\psp$ no significant signal was found~\cite{Bhardwaj:2011dj}. 
They measured the branching fractions
${\cal B}(B^\pm\to X(3872)K^\pm) {\cal B}(X(3872)\to
\jpsi\gamma) =(1.78^{+0.48}_{-0.44}\pm 0.12)\times 10^{-6}$, 
and provided upper limit on the  branching fraction 
${\cal B}(B^\pm\to X(3872)K^\pm)\cdot {\cal B}(X(3872)\to
\psp\gamma) <3.45 \times 10^{-6}$.
The upper limit on the ratio was
$\frac{{\cal B}(X(3872)\to \psp\gamma)}{{\cal B}(X(3872)\to
  \jpsi\gamma)}<2.1$ (at 90\% CL).

This results of $X(3872)\to \gamma \jpsi$ from Belle and BaBar are
consistent, while the $X(3872)\to \gamma \psp$ results are in
disagreement. More data is need to confirm these results.

\section*{The $B\to K\pi X(3872)$}
The production characteristic of the $X(3872)$ in $p\bar{p}$
collisions, such as the $p_T$ and rapidity distributions and the ratio
of prompt production versus production via $B$-meson decays, are very
similar to those of the well established $\psp$ charmonium state~\cite{Abazov:2004kp}. 
Thus it is of interest to compare production characteristics of the
$X(3872)$ to those of other charmonium states in $B$ meson decays. One
common characteristic of all of the known charmonium states that are
produced in $B$ meson decays is that when they are produced in
association with a $K\pi$ pair, the $K\pi$ system is always dominated
by a strong $K^*(890)\to K\pi$ signal. 

\begin{figure}[bpt]
  \includegraphics[width=70mm]{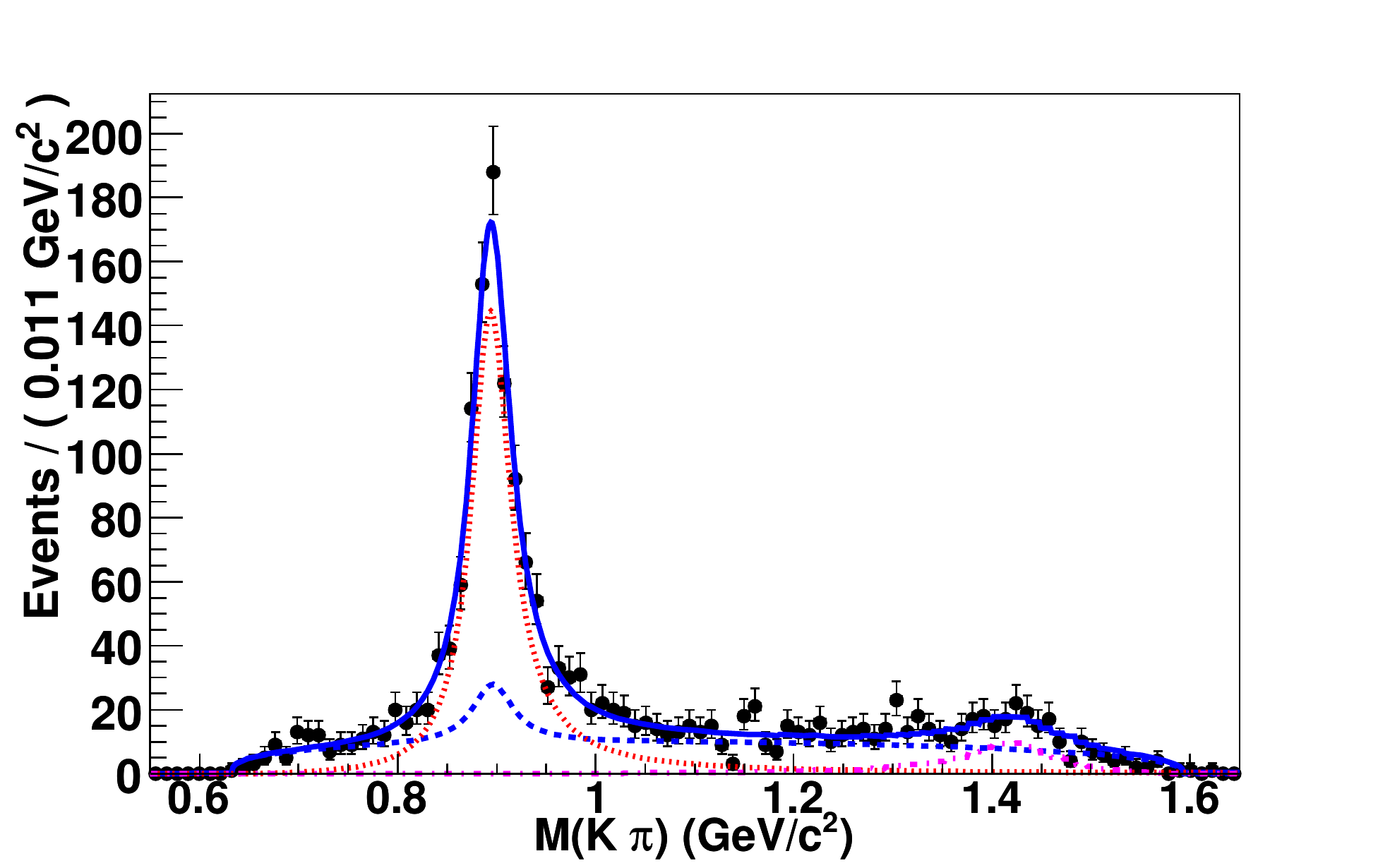}
  \caption{The $K\pi$ mass spectrum for the $B \to \psp K\pi$.  $B
    \to \psp K^{*}(892)^0$ is shown by the dotted red curve, $B \to
    \psp K^{*}_2(1430)^0$ by the dash-dot magenta curve, and the
    background by the dashed blue curve.}
  \label{fig:BtoPsipkpi}
\end{figure}

\begin{figure}[bpt]
  \includegraphics[width=70mm]{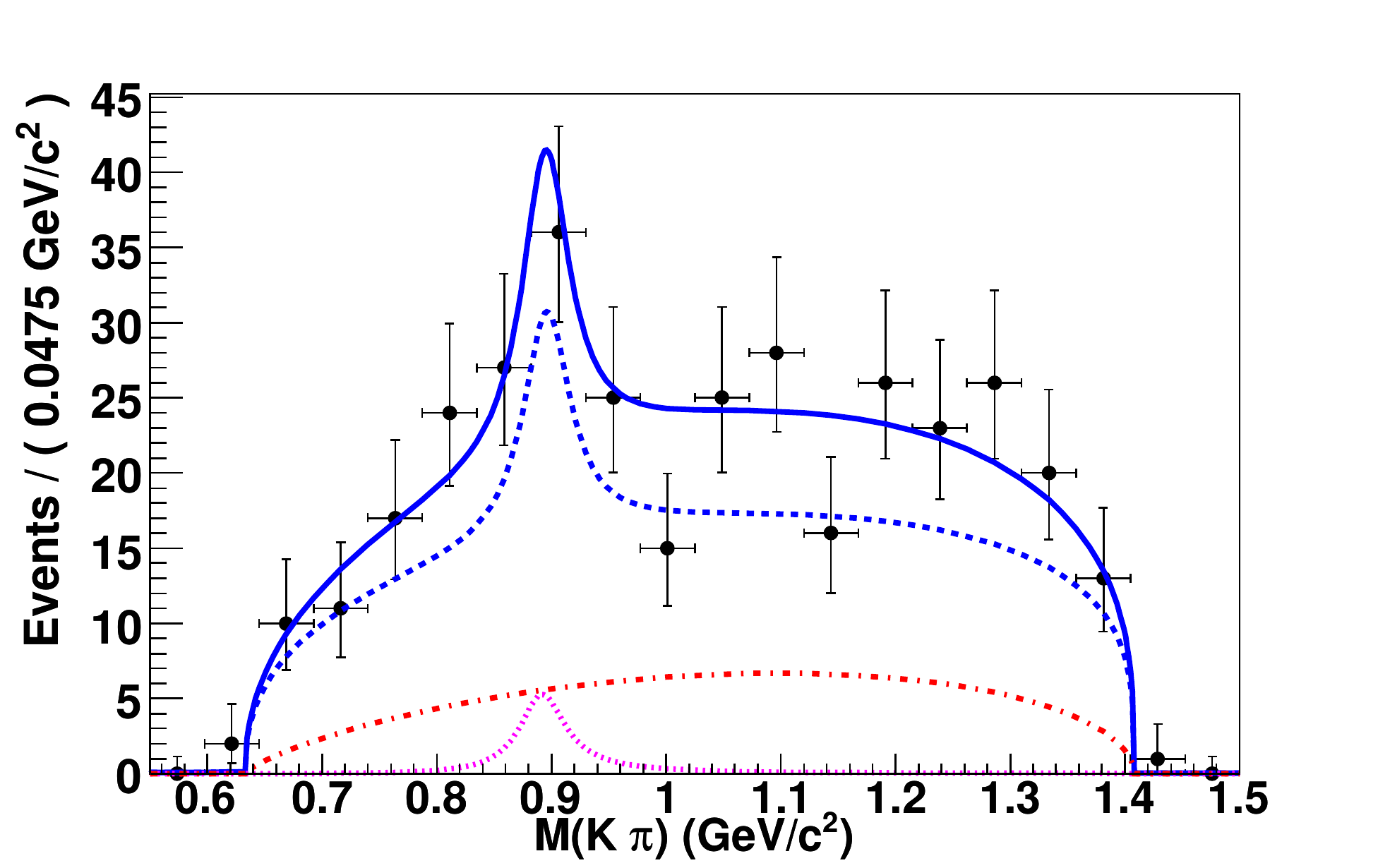}
  \caption{The $K\pi$ mass spectrum for the $B \to X(3872)K\pi$ candidates.
    $B \to X(3872) (K^+ \pi^-)_{NR}$ is shown by the dash-dot red curve,
    $B \to X(3872) K^{*}(892)^0$ by the dotted magenta curve,
    and the background by the dashed blue curve.
  }
  \label{fig:BtoXkpi}
\end{figure}

Belle studied the $X(3872)$ production in association with a $K\pi$ in
$B^0\to K^+\pi^-\jpsipp$ decays~\cite{Istomin:2008tj}. In a sample of 657M
$B\bar{B}$ pairs, about 90 $\jpsipp$ signal events are
seen. Figure~\ref{fig:BtoPsipkpi} shows the $K\pi$ invariant mass
distribution for these events, where it is evident that most of the
$K\pi$ pairs have a phase space like distribution, with little or no
signal for $K^*(890)\to K\pi$.
All of the events in the $K^*$ peak seem to be due to the side-band
determined background.
This is contrasted to the $B \to K \pi \psp$ events (with $\psp \to
\jpsipp$) in the data sample, where the $K\pi$ invariant
mass distribution, shown in Fig.~\ref{fig:BtoXkpi}, is dominated by
the $K^*(890)$.

The $X(3872)$ state remains a mystery.
Better understanding demands more experimental
constraints and theoretical insight.

\section*{The $Y$ states}
$Y(4260)$, the first unexpected vector charmonium-like state, was
observed by BaBar~\cite{Aubert:2005rm}
in ISR production of $Y(4260)\to \pi \pi \jpsi$.  CLEO~\cite{He:2006kg}
and Belle~\cite{:2007sj} confirmed the BaBar result, but Belle also
found an additional broader structure at 4008 $\mevcc$.
BaBar found~\cite{Aubert:2006ge} another  enhancement, $Y(4360)$ in
$\pp\psp$,  which Belle measured with larger mass and smaller
width, Belle also found~\cite{:2007ea} a second structure near
4660~$\mevcc$. 

\begin{figure}[bpt]
  \includegraphics[width=80mm]{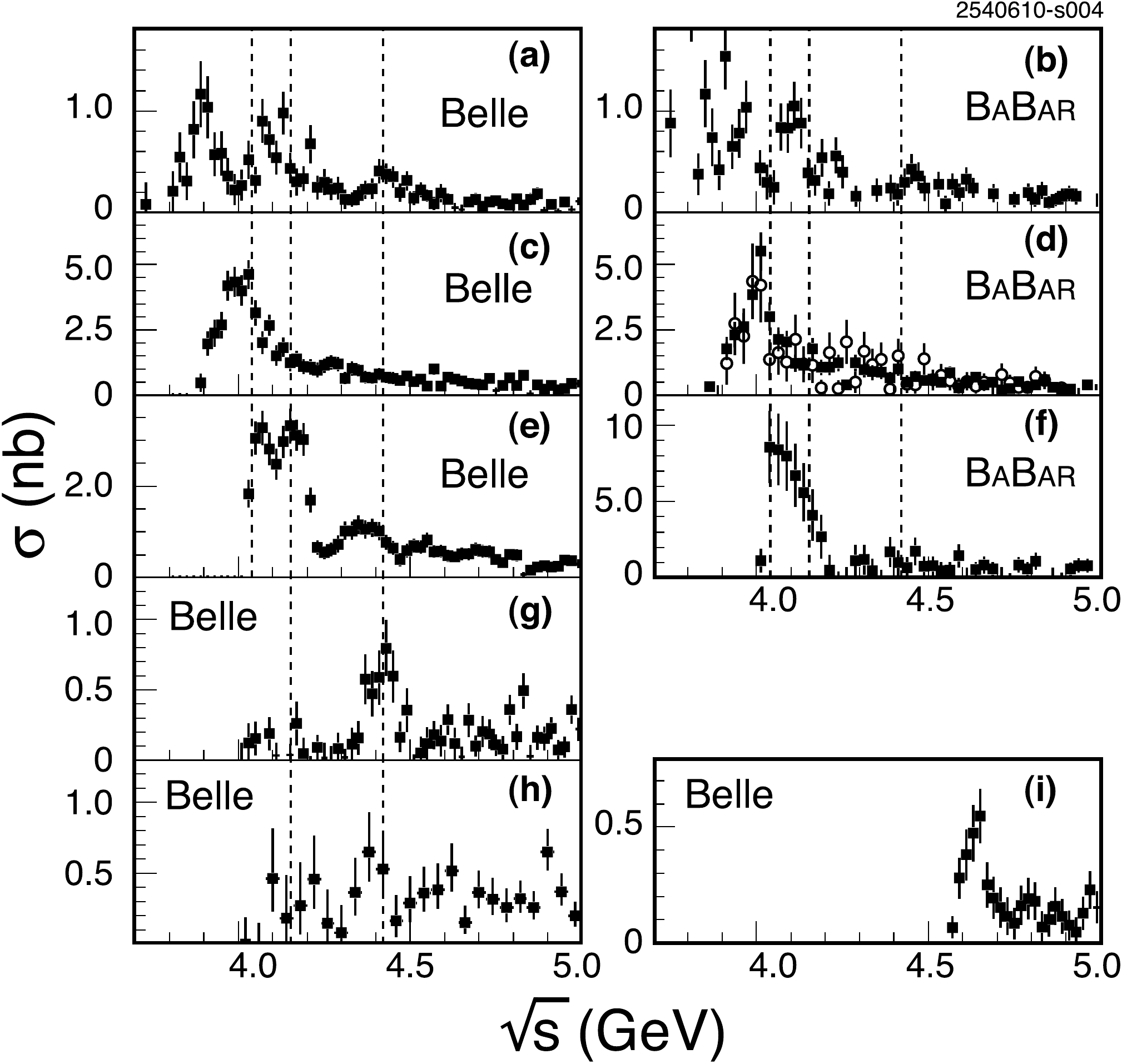}
  \caption{Measured $e^+e^-$ exclusive open-charm meson- or baryon-pair
    cross sections~\cite{Brambilla:2010cs},
    (a)~$D\bar{D}$ 
    (b)~$D\bar{D}$
    (c)~$D^+D^{*-}$
    (d)~$D\bar{D}^{*}$ for $D=D^0$ {\it (solid squares)}
    and $D=D^+$ {\it (open circles)} 
    (e)~$D^{*+}D^{*-}$ 
    (f)~$D^*D^*$ 
    (g)~$DD^+$ 
    (h)~$DD^{*+} $
    (i)~$\Lambda_c^+\Lambda_c^-$ 
  }
  \label{fig:DD-xsection}
\end{figure}

There is only one unassigned $1^{--}$ charmonium state in this mass region, the
$3^3D_1$, and no room to accommodate all of the four observed peaks.
Figure~\ref{fig:DD-xsection} shows the $D\bar{D}$ cross sections. No
enhancement  is seen for any $Y\to D^{(*)}D^{(*)}$. The absence of any evidence for
$Y(4260)$ ($Y(4360)$) decays to open charm implies that the
$\jpsipp$($\pp\psp$) partial width is large.
Ref.~\cite{Mo:2006ss} gives $\Gamma(Y(4260) \to \pi^+ \pi^- \jpsi)>508$~keV at 90\% CL,
an order of magnitude higher than expected for conventional vector
charmonium. Charmonium would also feature dominant
open charm decays, exceeding those of dipion transitions by a factor
expected to be $>100$, as is the case for the $\psi(3770)$ and
$\psi(4160)$.

At the EPS2011 conference, Belle presented an analysis of
$Y(4260)\to \piz\piz \jpsi$. They measured $\Gamma_{ee}{\cal B}(\jpsi \piz \piz)
=3.19^{+1.82+0.64}_{-1.53-0.35}$~ev, which is about half of the 
$\Gamma_{ee}{\cal B}(\jpsipp)=5.9^{+1.2}_{-0.9}$~ev~\cite{PDG}.
This is consistent with the CLEO's study~\cite{Coan:2006rv} on direct production of
$Y(4260)$ in $e^+e^-$ collisions.
It implies that the $Y(4260)$ has $I=0$, as expected for a $c\bar{c}$ state.

\subsection*{The decay $\psp\to\gamma\piz$, $\gamma\eta$ }
The study of vector charmonium decay to a photon and neutral
pseudoscalar meson $P=(\piz,\eta,\eta')$ provides experimental
constraints on the relevant QCD predictions, such as the vector meson
dominance mode(VDM), two-gluon couplings to $q\bar{q}$ states, mixing
of $\etac-\eta^{(')}$. 
The ratio $R_n \equiv {\cal B}(\psi(nS)\to \gamma \eta)/{\cal  B}(\psi(nS)\to \gamma \eta')$ 
is predicted by first order perturbation theory,
and $R_1 \approx R_2$  is also expected \cite{Chernyak:1983ej}.
However, CLEO-c reported measurements for the decays of $\jpsi$,
$\psp$ and $\psi''$ to $\gamma P$, and no evidence for
$\psp\to \gamma\eta$ or $\gamma\piz$ was found~\cite{:2009tia}. Therefore, they obtained
$R_2<<R_1$ with $R_2<1.8\%$ at 90\% CL. Such a small $R_2$ is
unanticipated, and it poses a significant challenge to our
understanding of the $c\bar{c}$ bound states.  Do other
processes contribute? Is this related to the $\rho\pi$ puzzle~\cite{Appelquist:1974zd}?

\begin{figure}[bpt]
  \includegraphics[width=60mm]{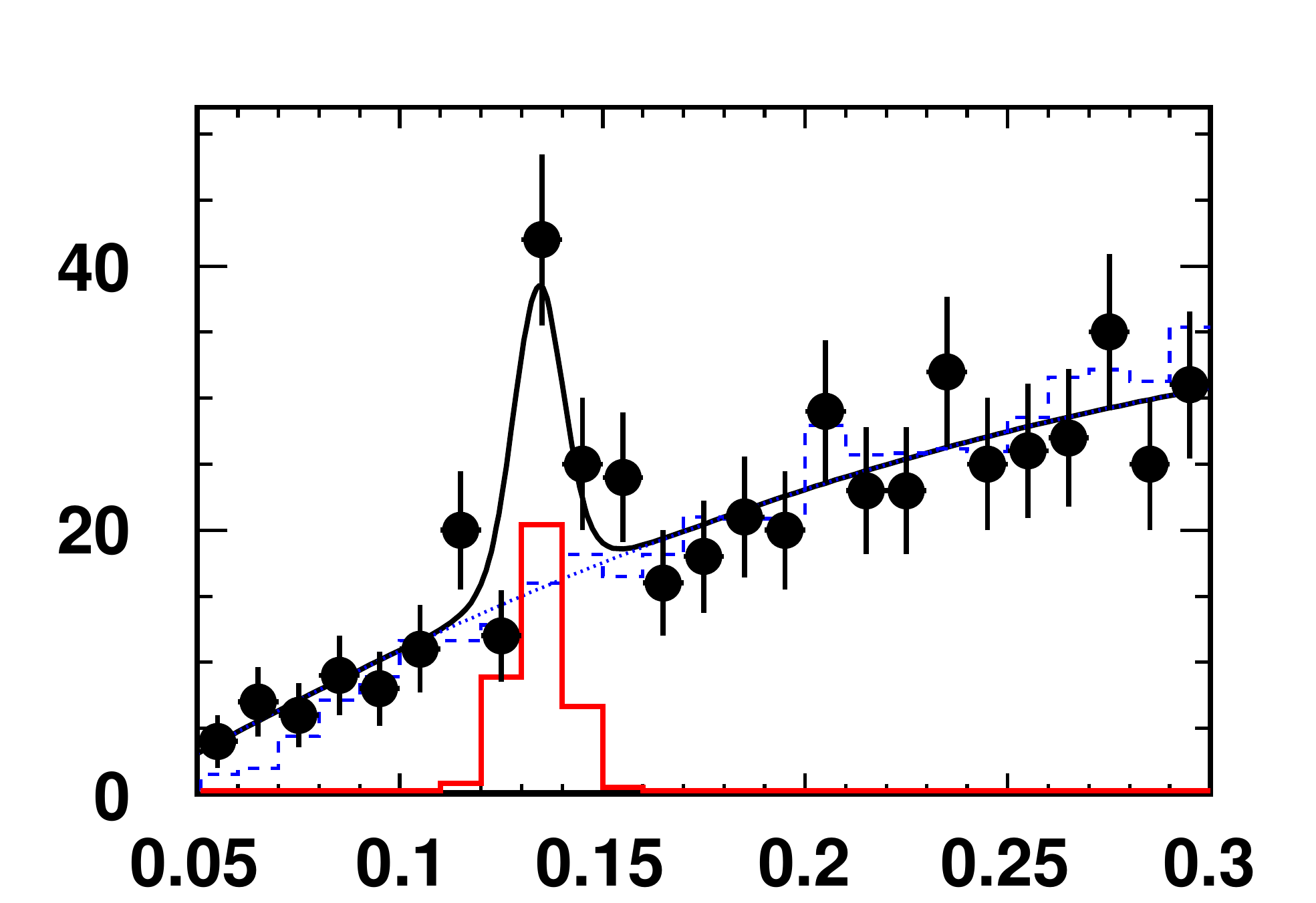}

  \includegraphics[width=40mm]{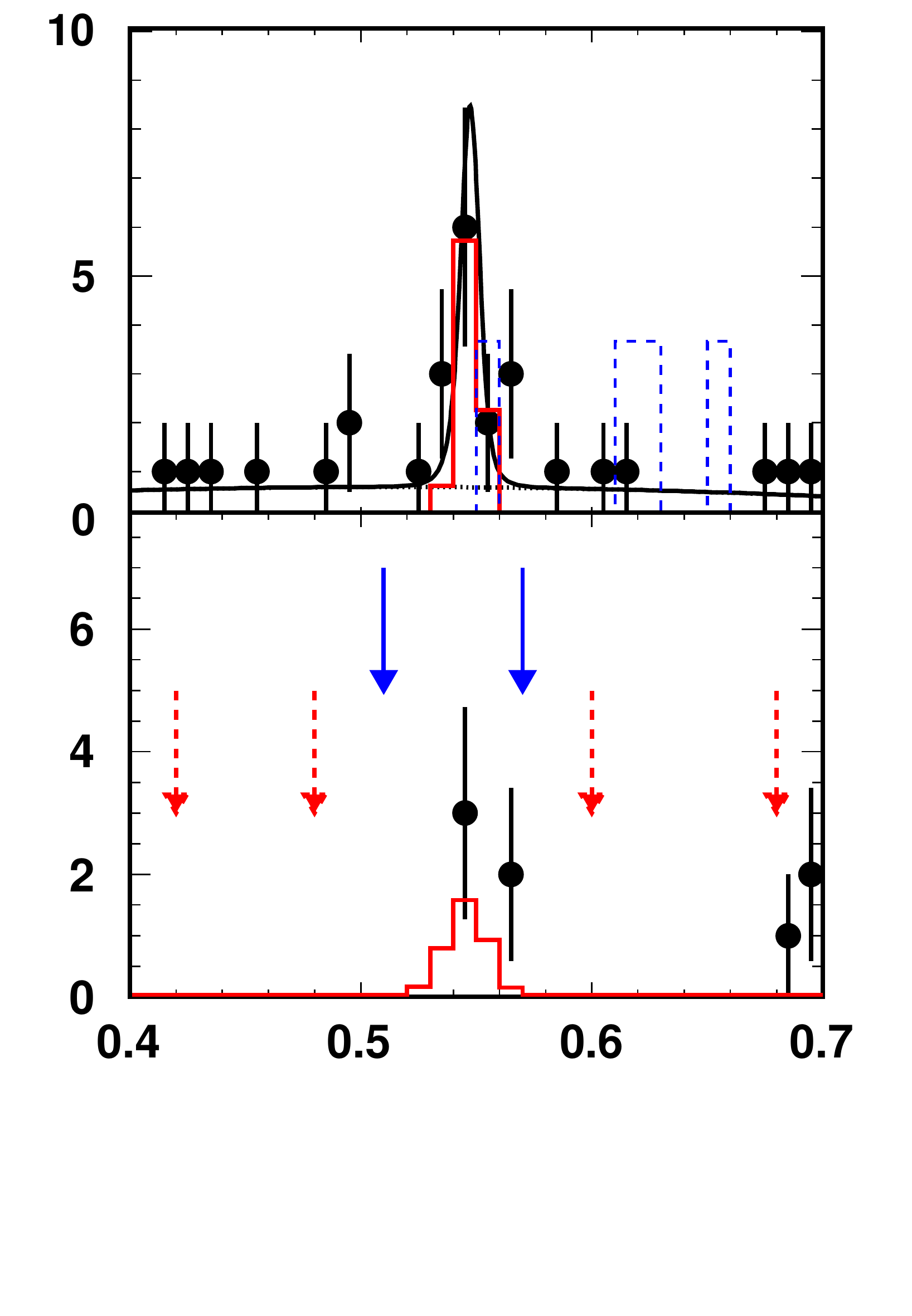}
  \includegraphics[width=40mm]{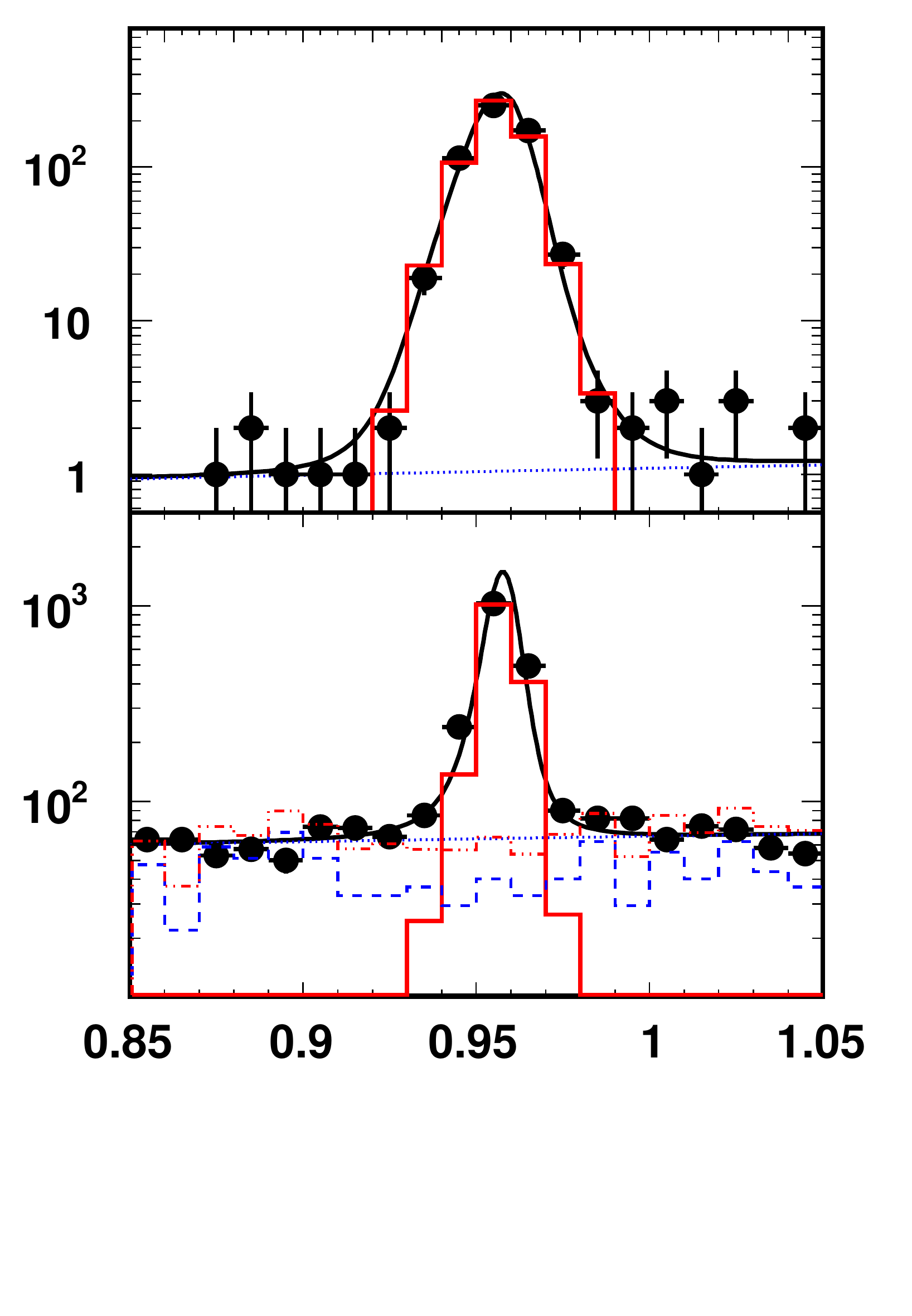}
    \put(-200,138){ \bf (b)}
    \put(-200,81){ \bf (c)}
    \put(-85, 138){ \bf (d)}
    \put(-85, 81){ \bf (e)}
  \caption{Mass distributions of the pseudoscalar meson candidates
  in $\psp \to \gamma P$: (a) $\gamma\piz$, (b) $\gamma\eta(\pip\pim\piz)$;
  (c) $\gamma\eta(3\piz)$; (d) $\gamma\etap[\pip\pim\eta(\gamma\gamma)]$; and
  (e) $\gamma\etap(\gamma\pip\pim)$. 
  }
  \label{fig:gammaP}
\end{figure}

With 106M $\psp$ events, BESIII observed $\psp\to \gamma\piz$ and
$\psp\to \gamma\eta$ and $\psp\to\gamma\etap$ , where $\eta$ is reconstructed from
$\eta\to \pp\piz$ and $\piz\piz\piz$, and $\etap$ is reconstructed from
$\etap\to \pp\eta$ and $\etap\to\gamma\pp$~\cite{Ablikim:2010dx}, as shown in Fig.~\ref{fig:gammaP}.
The measured branching fractions are summarize in Table.~\ref{tab:gammaP}.
The $R_2$ is about 20 times smaller than $R_1$. $Q\equiv
\frac{{\cal B}(\psp\to \gamma P)}{{\cal B}(\psp\to \gamma P)}$ for
each decay mode is also shown in the table, which is much smaller than 12\%.

\vspace*{-18cm}\hspace*{2cm} {\bf(a)}
\vspace*{18cm}\hspace*{-2cm}
\vspace*{-10.5cm} \hspace*{3cm} Mass (GeV/$c^2$)
\vspace*{10.5cm} \hspace*{-3cm}

\begin{table}[t]
\begin{center}
\caption{The measured branching fractions for $\psp\to
  \gamma\piz$, $\gamma\eta$, $\gamma\etap$. The branching fractions for
  $\jpsi$ decays are from the PDG.}
\begin{tabular}{|l|c|c|c|}\hline 
  mode & ${\cal B}(\psp) [\times 10^{-6}]$ &  ${\cal B}(\jpsi) [\times
    10^{-4}]$  & $ Q={\cal B}(\psp)/{\cal B}(\jpsi)$\\\hline 
$\gamma\piz$  & $1.58\pm 0.42$  & $0.35 \pm0.03$ &  $(4.5\pm  1.3)$\% \\ \hline 
$\gamma \eta$ & $1.38\pm 0.49$  & $11.04\pm0.34$ &  $(0.13\pm 0.04)$\% \\ \hline 
$\gamma\etap$ & $126\pm 9$      & $52.8 \pm 1.5$ &  $(2.4\pm 0.2)$\%   \\ \hline 
$R_{1/2}$     & $(1.10 \pm 0.39)$\% & $(20.9\pm0.9)$\% &  - \\ \hline
\end{tabular}
\label{tab:gammaP}
\end{center}
\end{table}

\section*{The decay $\chicJ \to \gamma V$, $VV$}
$\chicJ$ events make significant contributions to the radiative
decays of $\psp$. The decay of the P wave $\chicJ$ to $\gamma V$
provides a good chance to validate theoretical predictions and search
for glueballs~\cite{Amsler:1995td}. 
CLEO-c found~\cite{Bennett:2008aj} a surprisingly large $\chicJ\to \gamma V$
branching fraction, an order of magnitude higher than the pQCD prediction~\cite{Gao:2006bc}.
With 106M $\psp$ events, BESIII studied the decays $\chicJ \to \gamma
V$, with $V$ representing $\rho^0$, $\omega$ and
$\phi$~\cite{Ablikim:2011kv}. The results are listed in
Table~\ref{tab:gammaV}, where the decay $\chico\to \gamma\phi$ is the
first observation. The results provide tight constraints on QCD.

\begin{table}[t]
\begin{center}
  \caption{Compare the $\chicJ\to \gamma V$ branching fraction of pQCD calculations, with 
    measurements from CLEO-c and BESIII.}
\begin{tabular}{|l|c|c|c|}\hline 
  mode & pQCD  & CLEO-c & BESIII \\ \hline
$\chicz \to \gamma\rho^0$& 1.2  & $<9.6$           &$<10.5$ \\ 
$\chico \to \gamma\rho^0$& 14   & $243\pm 19\pm 22$&$228\pm13\pm16$ \\ 
$\chict \to \gamma\rho^0$& 4.4  & $<50$            &$<20.8$ \\ \hline 
$\chicz \to \gamma\omega$& 0.13 & $<8.8$           &$<12.9$ \\ 
$\chico \to \gamma\omega$& 1.6  & $83 \pm 15\pm 12$&$69.7\pm7.2\pm 5.6$ \\ 
$\chict \to \gamma\omega$& 0.5  & $<7.0$           &$<6.1$ \\ \hline
$\chicz \to \gamma\phi$  & 0.46 & $<6.4$           &$<16.2$ \\ 
$\chico \to \gamma\phi$  & 3.6  & $<26$            &$25.8\pm5.2\pm2.0$ \\ 
$\chict \to \gamma\phi$  & 1.1  & $<13$            &$<8.1$  \\ \hline
\end{tabular}
\label{tab:gammaV}
\end{center}
\end{table}

Vector pair decay modes are
measured at BESIII~\cite{:2011ih}. In the analysis, $\chicJ$ candidates
are reconstructed with $\phi\phi$, $\omega\omega$ and $\omega\phi$,
respectively. The helicity selection rule suppressed decays $\chico \to
\phi\phi/\omega\omega$ are observed for the first time and there is an
evidence of the doubly OZI suppressed decay $\chico \to \omega\phi$
with a significance of 4.1$\sigma$.

BESIII searched for $\etac(2S)\to VV$ decays using 106M $\psp$
events, where $VV$ represents $\rho^0\rho^0$, $K^{*0}K^{*0}$ and
$\phi\phi$~\cite{Collaboration:2011kr}. They found no evidence for $\etac(2S)\to VV$ signal, and
determined 90\% limits on the branching fractions, which are lower
than the theoretical predictions~\cite{Wang:2010iq}.

\section*{Summary}
Charmonium is the best understood hadronic system.
All the lowest-lying charmonium states have been found; the
long-anticipated states have been measured with high
precision, good agreement between their measured properties and
theory. Higher-mass charmonium meson searches have produced surprises;
unanticipated states showed up.

Enormous progress has been achieved on charmonium decays. Many
expected decays and transitions have either been measured with
high precision or for the first time. 

Belle, BaBar, CLEO, CDF and D0 have produced fruitful results in the past.
LHC is starting to produce physics results. Large data samples from LHC
will allow identification of the X, Y, Z states, and measurements of  production, and polarization.
BESIII will continue to study charmonium physics.
Future experiments, such as PANDA, will complement the activities at
other labs.

\bigskip 

\begin{acknowledgments}
This work is supported in part by the
Ministry of Science and Technology of China under Contract
No. 2009CB825200; 
\end{acknowledgments}

\bigskip 
\bibliography{basename of .bib file}

\end{document}